\documentclass[final, journal, twocolumn]{IEEEtran}

\usepackage[utf8]{inputenc}
\usepackage[british]{babel}

\usepackage{amsmath,amssymb,amsfonts}

\usepackage{siunitx}

\usepackage{eso-pic}

\usepackage{xr}

\allowdisplaybreaks

\newcommand{\CORRI}[1]{\textcolor{black}{#1}}
\newcommand{\CORRII}[1]{\textcolor{black}{#1}}

\newcommand{\FEATHER} [0] {Feather2$\ $}
\newcommand{\RTHINDEX} [0] {r$\HYPHEN$th$\ $}

\usepackage{textgreek}

\newcommand{\ITalpha}[0]{\alpha}

\newcommand{\ITkappa}[0]{\kappa}
\newcommand{\ITlambda}[0]{\lambda}
\newcommand{\ITmu}[0]{\mu}
\newcommand{\ITnu}[0]{\nu}
\newcommand{\ITxi}[0]{\xi}

\newcommand{\ITrho}[0]{\rho}

\newcommand{\ITsigma}[0]{\sigma}

\newcommand{\ITvarphi}[0]{\varphi}
\newcommand{\ITchi}[0]{\chi}

\newcommand{\ITomega}[0]{\omega}

\newcommand{\RMdelta}[0]{\textrm{\textdelta}}

\newcommand{\RMGamma}[0]{\Gamma}

\newcommand{\RMOmega}[0]{\Omega}

\usepackage{amsmath}

\newcommand{\HYPHEN} [0] {\operatorname{-}}
\newcommand{\RM}     [1] {\mathrm{#1}}
\newcommand{\BS}     [1] {\boldsymbol{#1}}
\newcommand{\BAR}    [1] {\overline{#1}}

\newcommand{\VEC}   [1] {\protect\vphantom{#1}\smash{\BS{\mathbf{#1}}}}
\newcommand{\VECl}  [2] {\protect\vphantom{#1}\smash{\BS{\mathbf{#1}}}_{#2}}
\newcommand{\VECu}  [2] {\protect\vphantom{#1}\smash{\BS{\mathbf{#1}}}^{#2}}
\newcommand{\VEClu} [3] {\protect\vphantom{#1}\smash{\BS{\mathbf{#1}}}_{#2}^{#3}}

\newcommand{\MAT}   [1] {\protect\vphantom{#1}\smash{\BS{\mathbf{#1}}}}

\newcommand{\MATu}  [2] {\protect\vphantom{#1}\smash{\BS{\mathbf{#1}}}^{#2}}

\newcommand{\SCA}   [1] {\protect\vphantom{#1}\smash{\RM{#1}}}
\newcommand{\SCAl}  [2] {\protect\vphantom{#1}\smash{\RM{#1}}_{#2}}
\newcommand{\SCAu}  [2] {\protect\vphantom{#1}\smash{\RM{#1}}^{#2}}
\newcommand{\SCAlu} [3] {\protect\vphantom{#1}\smash{\RM{#1}}_{#2}^{#3}}

\newcommand{\GRAD} [0] {\nabla}
\newcommand{\DIV}  [0] {\nabla\cdot}
\newcommand{\CURL} [0] {\nabla\times}

\newcommand{\GRADTANG} [0] {\SCAl{\nabla}{\RM{t}}}

\newcommand{\CURLTANG} [0] {\SCAl{\nabla}{\RM{t}}\times}

\newcommand{\GRADNORM} [0] {\SCAl{\nabla}{\RM{n}}}

\newcommand{\D}     [1] {\!\mathop{}\mathrm{d}{#1}}
\newcommand{\dxP}   [1] {\partial_{#1}}
\newcommand{\dx}    [1] {\!\mathrm{\frac{d}{d{#1}}}}

\newcommand{\TRANS} [0] {\mkern 1mu\top}

\usepackage{pgfplots}\pgfplotsset{compat=1.14}
\usepackage{PGFplot/PGFplotTUDA/pgfplots-private}
\usepgfplotslibrary{units} 

\newcommand{\FigureCurrentSharingFunction}[0]{
    \begin{tikzpicture}
		\begin{axis}[	
            temflineplot,
            legend pos  = north west,
            width       = 8cm, 
            height      = 6cm,
            xmin        = 0, 
            xmax        = 1,
            ymin        = -1, 
            ymax        = 1,	
            xlabel      = $\SCAu{\ITlambda}{r}$, 
            ylabel      = $f(\SCAu{\ITlambda}{r})$,
            x unit      = \HYPHEN, 
            y unit      = \HYPHEN,
            ]
            
            \addplot+ 
                [no marks] 
                table[x=lambda,y=Ic, col sep=comma] 
                {Data/Results_CurrentSharingFunction.csv}; 
            \addplot+ 
                [no marks] 
                table[x=lambda,y=2Ic, col sep=comma] 
                {Data/Results_CurrentSharingFunction.csv};  
            \addplot+ 
                [no marks] 
                table[x=lambda,y=5Ic,col sep=comma] 
                {Data/Results_CurrentSharingFunction.csv};     
            \addplot+ 
                [no marks] 
                table[x=lambda,y=20Ic,col sep=comma] 
                {Data/Results_CurrentSharingFunction.csv};
            
            \node
                [anchor=west] at (0.1, 0.7)
                {\footnotesize{$\SCAu{k}{r}=20$}};  
            \node
                [anchor=west] at (0.4, 0.7)
                {\footnotesize{$\SCAu{k}{r}=5$}}; 
            \node
                [anchor=west] at (0.75, 0.7)
                {\footnotesize{$\SCAu{k}{r}=2$}}; 
            \node
                [anchor=west] at (0.75, -0.5)
                {\footnotesize{$\SCAu{k}{r}=1$}};    
      
	    \end{axis}
	\end{tikzpicture}
}

\newcommand{\TapeTwoDCurrentDensity}[0]{
    \begin{tikzpicture}
		\begin{semilogyaxis}[	
            temflineplot,
            legend pos  = north west,
            width       = 8cm, 
            height      = 6cm,
            xmin        = 0, 
            xmax        = 2.5,
            ymin        = 1e-6, 
            ymax        = 1e+2,	
            ytickten    = {-6,...,+2},
            xlabel      = $\RM{Normalized\ current}$, 
            ylabel      = $\RM{loss/cycle}$,
            x unit      = \HYPHEN, 
            y unit      = \si{\joule},
            ]
            
            \addplot+ 
                [only marks] 
                table[x=IIc,y=H1Hz,col sep=comma]
                {Data/Comparison_ACloss_f.csv}; 
            \addplot+ 
                [only marks] 
                table[x=IIc,y=H10Hz,col sep=comma]
                {Data/Comparison_ACloss_f.csv};  
            \addplot+ 
                [only marks] 
                table[x=IIc,y=H100Hz,col sep=comma]
                {Data/Comparison_ACloss_f.csv}; 
            \addplot+ 
                [only marks] 
                table[x=IIc,y=H1000Hz,col sep=comma]
                {Data/Comparison_ACloss_f.csv};
                        
            \pgfplotsset{cycle list shift=-4} 
            
            \addplot+ 
                [mark=none] 
                table[x=IIc,y=HA1Hz,col sep=comma]
                {Data/Comparison_ACloss_f.csv}; 
            \addplot+ 
                [mark=none] 
                table[x=IIc,y=HA10Hz,col sep=comma]
                {Data/Comparison_ACloss_f.csv};       
            \addplot+ 
                [mark=none] 
                table[x=IIc,y=HA100Hz,col sep=comma]
                {Data/Comparison_ACloss_f.csv};      
            \addplot+ 
                [mark=none] 
                table[x=IIc,y=HA1000Hz,col sep=comma]
                {Data/Comparison_ACloss_f.csv};
            
            \node
                [anchor=west] at (2, 1e+1) 
                {\footnotesize{$\SI{1}{\hertz}$}};  
            \node
                [anchor=west] at (2, 1e+0) 
                {\footnotesize{$\SI{10}{\hertz}$}}; 
            \node
                [anchor=west] at (2, 1e-1) 
                {\footnotesize{$\SI{100}{\hertz}$}}; 
            \node
                [anchor=west] at (2, 1e-2) 
                {\footnotesize{$\SI{1000}{\hertz}$}};     
            
            \filldraw 
                [fill=white, draw=black] (0.1, 1.2e0) 
                rectangle ++(1.2,40);            
            \node
                [anchor=west] at (0.15, 1.5e1) 
                {\footnotesize{Marker: H (reference)}};
            \node
                [anchor=west] at (0.15, 3e0) 
                {\footnotesize{Line: A-H (thin shell)}};      
      
	    \end{semilogyaxis}
	\end{tikzpicture}
}

\newcommand{\FigComparisonComputationalTime}[0]{
    \begin{tikzpicture}
		\begin{loglogaxis}[	
            temflineplot,
            legend pos  = north west,
            width       = 8cm, 
            height      = 6cm,
            xmin        = 1e0, 
            xmax        = 1e4,
            ymin        = 1e-3, 
            ymax        = 1e+3,	
            xtickten    = {0,...,+4},
            ytickten    = {-3,...,+3},
            xlabel      = $\RM{Number\ of\ tapes}$, 
            ylabel      = $\RM{Time}$,
            x unit      = \HYPHEN, 
            y unit      = \si{\hour},
            ]
            
            \addplot+ 
                table[x=x,y=H,col sep=comma] 
                {Data/Comparison_computationalTime.csv}; 
            
            \pgfplotsset{cycle list shift=-1}
            
            \addplot+ 
                [mark=none, dashed] 
                table[x=x,y=H-extrap,col sep=comma] 
                {Data/Comparison_computationalTime.csv}; 
            
            \pgfplotsset{cycle list shift=-1}
            
            \addplot+ 
                table[x=x,y=HA,col sep=comma]
                {Data/Comparison_computationalTime.csv}; 
				
	        \legend{$\RM{H: reference}$,
		            $\RM{H: extrapolation}$,
			        $\RM{A}$-$\RM{H: thin\ shell}$,	
			        }   
      
	    \end{loglogaxis}
	\end{tikzpicture}
}

\newcommand{\FigPoweringFeather}[0]{
    \begin{tikzpicture}
		\begin{axis}[	
            temflineplot,
            legend pos  = north west,
            width       = 7cm, 
            height      = 6cm,
            xmin        = -400, 
            xmax        = 200,
            ymin        = -40, 
            ymax        = 40,	
            xlabel      = $\RM{Time}$, 
            ylabel      = $\RM{Voltage}$,
            x unit      = \si{\second}, 
            y unit      = \si{\milli\volt},
            ]
            
            \addplot+ 
                [no marks, dashed] 
                table[x=s,y=mV,col sep=comma] 
                {Data/Powering_FeatherM2.csv};
            \label{plot_1_y1}
            
        \end{axis} 
       
      \begin{axis}[	
		    temflineplot,
		    axis y line*= right,
		    xmajorgrids = false,
	    	ymajorgrids = false,
            legend pos  = south east,
            width       = 7cm, 
            height      = 6cm,
            xmin        = -400, 
            xmax        = 200,
            ymin        = -10, 
            ymax        = 10,	
            xlabel      = ,
            ylabel      = $\RM{Current}$,
            x unit      = , 
            y unit      = \si{\kilo\ampere},
            xticklabels = {,,}
            ]  
              
            \pgfplotsset{cycle list shift=+1}                          
            
            \addplot+ 
                [no marks] 
                table[x=s,y=A,col sep=comma]
                {Data/Powering_FeatherM2.csv};            
            \label{plot_1_y2}
		
	        \addlegendimage{/pgfplots/refstyle=plot_1_y1}
	        \addlegendentry{$\RM{Current}$}
            \addlegendimage{/pgfplots/refstyle=plot_1_y2}
            \addlegendentry{$\RM{Voltage}$}	
				
	    \end{axis}			
	\end{tikzpicture}
}

\begin{document}


\AddToShipoutPicture*{
    \footnotesize\sffamily\raisebox{0.8cm}{\hspace{1.5cm}\fbox{
        \parbox{\textwidth}{
            \copyright~2019
                IEEE. Personal use of this material is permitted. Permission from IEEE
                must be obtained for all other uses, in any current or future media, 
                including reprinting/republishing this material for advertising or
                promotional purposes, creating new collective works, for resale or
                redistribution to servers or lists, or reuse of any copyrighted
                component of this work in other works.
            }
        }
    }
}


\title{A Coupled A-H Formulation for Magneto-Thermal Transients in High-Temperature Superconducting Magnets}

	\author{
            {
			L. Bortot, 
            B. Auchmann,
            I. Cortes Garcia,            
            H. De Gersem,
			M. Maciejewski,  
			M. Mentink,
			S. Sch{\"o}ps,
		}
		
			{
			J. Van Nugteren,
			and A.P. Verweij
		}
		
		\thanks{
        
            L. Bortot is with CERN, Geneva, Switzerland, and with Technische Universit{\"a}t Darmstadt, Darmstadt, Germany (e-mail: lorenzo.bortot@cern.ch).
            
            I.C. Garcia, H. De Gersem and S. Sch{\"o}ps are with  Technische Universit{\"a}t Darmstadt, Darmstadt, Germany.    
            
            B.Auchmann is with CERN, Switzerland, and with Paul Scherrer Institute, Villigen, Switzerland.
            
            M. Maciejewski, M. Mentink, J. Van Nugteren and A.P. Verweij are with CERN, Geneva, Switzerland.
            
            This work has been sponsored by the Wolfgang Gentner Programme of the German Federal Ministry of Education and Research (grant no. 05E15CHA), by the ‘Excellence Initiative’ of the German Federal and State Governments and by the Graduate School of Computational Engineering at Technische Universit{\"a}t Darmstadt. Parts of the work have been funded by the BMBF project 05P18RDRB1 “Quench simulation for superconducting magnets: increase of the resolution in time and space” within the collaboration “Diagnose of high-intensity hadron beams (DIAGNOSE)”.
        }
	}


\maketitle
\IEEEpeerreviewmaketitle



\begin{abstract}

    The application of high-temperature superconductors to accelerator magnets for future particle colliders is under study. Numerical methods are crucial for an accurate evaluation of the complex dynamical behavior of the magnets, especially concerning the magnetic field quality and thermal behavior. We present a coupled $\VEC{A}\HYPHEN\VEC{H}$ field formulation for the analysis of magneto-thermal transients in accelerator magnets. The magnetic field strength $\VEC{H}$ accounts for the eddy current problem in the source regions containing the superconducting domains, while the magnetic vector potential $\VEC{A}$ represents the magnetoquasistatic problem in the normal and non-conducting domains. Furthermore, we include a \CORRI{thin shell} approximation for the source regions, making the formulation suitable for large scale models composed of thousands of tapes. In this work, the relevant equations are derived and discussed, with emphasis on the coupling conditions. The weak formulation is derived, and numerical results are provided in order to both, verify the formulation and scale it to the size of an accelerator magnet.
    
\end{abstract}
 
 






\newcommand{\EQdefJpowerLaw}
   { 
   \SCA{\ITrho}(|\VEC{J}|) =
   \frac{\SCAl{E}{\RM{crit}}}{\SCAl{J}{\RM{crit}}}
   \left(\frac{|\VEC{J}|}{\SCAl{J}{\RM{crit}}}\right)^{\SCA{n}-1}
   }
   
\newcommand{\EQdefUs}  
    {
    \VECl{u}{\RM{s}} =
    [\SCAu{u}{1},\ldots,
     \SCAu{u}{\SCAl{N}{\RM{r}}}]^{\TRANS}
    }
    
\newcommand{\EQdefIs}  
    {
    \VECl{i}{\RM{s}} =
    [\SCAu{i}{1},\ldots,
     \SCAu{i}{\SCAl{N}{\RM{r}}}]^{\TRANS}
    }
    
\newcommand{\EQdefCHI}  
    {
    \VEC{\ITchi} =
    [\VECu{\ITchi}{1},\ldots,
     \VECu{\ITchi}{\SCAl{N}{\RM{r}}}]^{\TRANS}
    } 

\newcommand{\EQdefE}
    {
    \VEC{E} =
    \SCA{\ITrho}\CURL\VEC{H}+\VECl{E}{\RM{s}}
    }

\newcommand{\EQdefEr}  
    {
    \VECu{E}{r} =
    \SCA{\ITrho}\CURL\VECu{H}{r}\CORRII{-}\VECu{\ITchi}{r}\SCAu{u}{r}
    }

\newcommand{\EQdefEs}  
    {
    \VECl{E}{\RM{s}} =
    \CORRII{-}\sum\nolimits_{r=1}^{\SCAl{N}{\RM{r}}}\VECu{\ITchi}{r}\SCAu{u}{r}
    }
    


\newcommand{\EQstrongAampereMaxwell}
  { 
  \CURL\ITmu^{-1}\CURL\VECu{A}{\star} 
  + \SCA{\ITsigma}\dxP{t}\VECu{A}{\star} 
  & = 0
  }
   
\newcommand{\EQstrongHfaraday}
  { 
  \CURL\SCA{\ITrho}\CURL\VECu{H}{r} 
  + \dxP{t}\SCA{\ITmu}\VECu{H}{r} 
  \CORRII{-} \CURL\VECu{\ITchi}{r}\SCAu{u}{r} 
  & = 0
  }
   
\newcommand{\EQconstraintIsource}
  { 
  \int\limits_{\SCAlu{\RMOmega}{\RM{H}}{r}}\!\!
  \VECu{\ITchi}{r}\cdot(\CURL\VECu{H}{r})
  \D{\SCA{\RMOmega}} 
  }

\newcommand{\EQdirichletAHtangA}
    { 
    \VECu{A}{\star}\times\VECl{n}{\RMOmega} 
    & = 0
    } 
    
\newcommand{\EQdirichletAHtangE}
    { 
    \VECu{E}{r}\times\CORRII{\VEClu{n}{\RMOmega}{r}}
    & = 0
    } 

\newcommand{\EQinterfaceAHnormJ}
    { 
    (\SCA{\ITsigma}\dxP{t}\VECu{A}{\star}+\CURL\VECu{H}{r})
    \cdot\CORRII{\VEClu{n}{\RMOmega}{r}}
    & = 0
    } 

\newcommand{\EQinterfaceAHnormB}
    { 
    (\CURL\VECu{A}{\star}-\SCA{\ITmu}\VECu{H}{r})
    \cdot\CORRII{\VEClu{n}{\RMOmega}{r}}
    & = 0
    } 

\newcommand{\EQinterfaceAHtangH}
    { 
    (\ITmu^{-1}\CURL\VECu{A}{\star}-\VECu{H}{r})
    \times\CORRII{\VEClu{n}{\RMOmega}{r}}
    & = 0
    }
  
\newcommand{\EQinterfaceAHtangE}
    { 
    (\dxP{t}\VECu{A}{\star} 
    + \SCA{\ITrho}\CURL\VECu{H}{r} 
    \CORRII{-} \VECu{\ITchi}{r}\SCAu{u}{r})
    \times\CORRII{\VEClu{n}{\RMOmega}{r}} 
    & = 0
    } 

\newcommand{\EQdefJjouleLoss}
  { 
  \SCAl{P}{\RM{Joule}}
  = \SCA{\ITrho}|\VEC{J}|^{2}
  }

\newcommand{\EQstrongTheatBalance}
    {    
    \SCAl{\ITrho}{\RM{m}}\SCAl{C}{\RM{p}}\dxP{t}\SCA{T}
    - \DIV(\SCA{\ITkappa}\GRAD\SCA{T}) 
    & = \SCAl{P}{\RM{Joule}}
    } 


\newcommand{\EQweakAampereMaxwell}
    {
    & \int\limits_{\SCAl{\RMOmega}{\RM{A}}}
    (\ITmu^{-1}\CURL\VECu{A}{\star})\cdot\CURL\VECl{w}{i}
    \D{\SCA{\RMOmega}}
    + \int\limits_{\SCAl{\RMOmega}{\RM{A}}}
    (\ITsigma\dxP{t}\VECu{A}{\star})\cdot\VECl{w}{i}
    \D{\SCA{\RMOmega}} \\
    & - \int\limits_{\SCAlu{\RMGamma}{\RM{HA}}{r}}\!\!
    \left[(\ITmu^{-1}\CURL\VECu{A}{\star})\times\VECl{w}{i}\right]\cdot
    \D{\VEC{\SCA{\RMGamma}}{}{}}
    =0
    }
    
\newcommand{\EQweakHfaraday}
    {
    & \int\limits_{\SCAlu{\RMOmega}{\RM{H}}{r}}
    (\SCA{\ITrho}\CURL\VECu{H}{r})\cdot\CURL\VEClu{v}{p}{r}
    \D{\SCA{\RMOmega}}
    + \int\limits_{\SCAlu{\RMOmega}{\RM{H}}{r}}
    (\dxP{t}\SCA{\ITmu}\VECu{H}{r})\cdot\VEClu{v}{p}{r}
    \D{\SCA{\RMOmega}} \\
    & \CORRII{-} \int\limits_{\SCAlu{\RMOmega}{\RM{H}}{r}}
    \VECu{\ITchi}{r}\SCAu{u}{r}\cdot\CURL\VEClu{v}{p}{r}
    \D{\SCA{\RMOmega}}
    + \int\limits_{\SCAlu{\RMGamma}{\RM{HA}}{r}}\!\!
    (\VECu{E}{r}\times\VEClu{v}{p}{r})\cdot
    \D{\VEC{\SCA{\RMGamma}}{}{}}
    = 0
    }
    
 \newcommand{\EQweakTheatBalance}
    {
    & \int\limits_{\SCA{\RMOmega}}
    (\SCA{\ITkappa}\GRAD\SCA{T})\cdot\GRAD\SCAl{N}{m}
    \D{\SCA{\RMOmega}}
    + \int\limits_{\SCA{\RMOmega}} 
    (\SCAl{\ITrho}{\RM{m}}\SCAl{C}{\RM{p}}\dxP{t}\SCA{T})\SCAl{N}{m}
    \D{\SCA{\RMOmega}} \\
    & = \int\limits_{\SCA{\RMOmega}} 
   \SCAl{P}{\RM{Joule}}\SCAl{N}{m}
    \D{\SCA{\RMOmega}}
    }   

\newcommand{\EQconstraintIsourcePort}
  { 
  - \int\limits_{\SCAlu{\RMGamma}{\RM{J}}{r}}\!\!
  (\CURL\VECu{H}{r})\cdot
  \D{\SCA{\RMGamma}}
  }
  

\newcommand{\EQdefAdiscrete}
    {
    \VECu{A}{\star}=\sum_{j}\VECl{w}{j}\SCAl{a}{j}
    }
    
\newcommand{\EQdefHdiscrete}
    {
    \VECu{H}{r} =
    \sum_{q}\VEClu{v}{q}{r}\SCAlu{h}{q}{r}
    }
    
\newcommand{\EQdefCHIdiscrete}
    {
    \VECu{\ITchi}{r} =
    -\sum_{p}\GRAD\SCAlu{N}{p}{r}\SCAlu{\ITxi}{p}{r} =   
    \sum_{p}\VEClu{x}{p}{r}
    }
    
\newcommand{\EQdefTdiscrete}
    {
    \SCA{T}=\sum_{n}\SCAl{N}{n}\SCAl{t}{n}
    }
    
\newcommand{\EQmatAHTdiscrete}
    {
    \newcommand{\KA}{\MATu{K}{\ITnu}\!\!+\!\MATu{M}{\ITsigma}\dx{t}}
    \newcommand{\Q}{-\MAT{Q}}
    \newcommand{\QT}{\MATu{Q}{\TRANS}\dx{t}}   
    \newcommand{\M}{\MATu{K}{\ITrho}\!\!+\!\MATu{M}{\ITmu}\dx{t}}
    \newcommand{\X}{-\MAT{X}}
    \newcommand{\XT}{\MATu{X}{\TRANS}}
    \newcommand{\NULL}{\MAT{0}}   
    \newcommand{\T}{\MATu{K}{\ITkappa}\!\!+\!\MATu{M}{\ITrho}\dx{t}}    
    \renewcommand*{\arraystretch}{1.25}        
    \begin{bmatrix} 
        \!\KA   & \!\!\!\!\Q    & \!\!\!\!\NULL & \!\!\!\!\NULL \!\\ 
        \!\QT   & \!\!\!\!\M    & \!\!\!\!\X    & \!\!\!\!\NULL \!\\ 
        \!\NULL & \!\!\!\!\XT   & \!\!\!\!\NULL & \!\!\!\!\NULL \!\\ 
        \!\NULL & \!\!\!\!\NULL & \!\!\!\!\NULL & \!\!\!\!\T    \!\\[3pt]     
    \end{bmatrix} 
    \!\!\cdot\!\!
    \begin{bmatrix}  
         \VEC{a}  \\
         \VEC{h}  \\
         \VECl{u}{\RM{s}}  \\
         \VEC{t}  \\[3pt]
    \end{bmatrix}  
    \!\!=\!\!
    \begin{bmatrix}   
         \! \NULL          \!\! \\
         \! \NULL          \!\! \\
         \! \VECl{i}{\RM{s}}          \!\! \\
         \! \VEC{q}(\cdot) \! \\[3pt]
    \end{bmatrix}
    }
        
\newcommand{\EQdiscreteCoeffAHlabelKnu}
    {
    (\SCAu{K}{\ITnu})_{i,j} &=
    \int\limits_{\SCAl{\RMOmega}{\RM{A}}}
    (\ITmu^{-1}\CURL\VECl{w}{j}) \cdot\CURL\VECl{w}{i}
    \D{\SCA{\RMOmega}}
    }

\newcommand{\EQdiscreteCoeffAHlabelMsigma}
    {
    (\SCAu{M}{\ITsigma})_{i,j} &=
    \int\limits_{\SCAl{\RMOmega}{\RM{A}}}
    (\ITsigma\VECl{w}{j}) \cdot\VECl{w}{i}
    \D{\SCA{\RMOmega}}
    }

\newcommand{\EQdiscreteCoeffAHlabelQ}
    {
    (\SCA{Q})_{i,q}^{r} &=
    \int\limits_{\SCAlu{\RMGamma}{\RM{HA}}{r}}
    (\VEClu{v}{q}{r}\times\VECl{w}{i})\cdot
    \D{\VEC{\SCA{\RMGamma}}{}{}}
    }

\newcommand{\EQdiscreteCoeffAHlabelKrho}
    {
    (\SCAu{K}{\ITrho})_{p,q}^{r} &=
    \int\limits_{\SCAlu{\RMOmega}{\RM{H}}{r}}
    (\SCA{\ITrho}\CURL\VEClu{v}{q}{r})\cdot\CURL\VEClu{v}{p}{r}
    \D{\SCA{\RMOmega}}
    }

\newcommand{\EQdiscreteCoeffAHlabelMmu}
    {
    (\SCAu{M}{\ITmu})_{p,q}^{r} &=
    \int\limits_{\SCAlu{\RMOmega}{\RM{H}}{r}}
    (\SCA{\ITmu}\VEClu{v}{q}{r})\cdot\VEClu{v}{p}{r}
    \D{\SCA{\RMOmega}}
    }
    
\newcommand{\EQdiscreteCoeffAHlabelCHIOLD}
    {
    (\SCA{X})_{p}^{r} &=
    -\int\limits_{\SCAlu{\RMGamma}{\RM{J}}{r}}
    (\CURL\VEClu{v}{p}{r})\cdot
    \D{\VEC{\RMGamma}}
    }
    
\newcommand{\EQdiscreteCoeffAHlabelCHI}
    {
    (\SCA{X})_{p}^{r} &=
    \CORRII{\int\limits_{\SCAlu{\RMOmega}{\RM{H}}{r}}
    \VEClu{x}{p}{r}\cdot
    (\CURL\VEClu{v}{p}{r})
    \D{\SCA{\RMOmega}}}
    }
    
\newcommand{\EQdiscreteCoeffTlabelKk}
    {
    (\SCAu{K}{\ITkappa})_{m,n}^{} &=
    \int\limits_{\SCA{\RMOmega}}
    (\ITkappa\GRAD\SCAl{N}{n})\cdot\GRAD\SCAl{N}{m}
    \D{\SCA{\RMOmega}}
    }

\newcommand{\EQdiscreteCoeffTlabelMrho}
    {
    (\SCAu{M}{\ITrho})_{m,n}^{} &=
    \int\limits_{\SCA{\RMOmega}}
    (\SCAl{\ITrho}{\RM{m}}\SCAl{C}{\RM{p}}\SCAl{N}{n})\SCAl{N}{m}
    \D{\SCA{\RMOmega}}
    }

\newcommand{\EQdiscreteCoeffTlabelPjoule}
    {
    (\SCA{q}(\cdot))_{m} &=
    \int\limits_{\SCA{\RMOmega}}
    \SCA{q}(\cdot)\SCAl{N}{m}
    \D{\SCA{\RMOmega}}
    }


\newcommand{\EQdefAHimpedanceZ}  
    {
    \MAT{Z}(j\ITomega) &=
    [\MATu{X}{\TRANS}
    [\MATu{K}{\ITrho}+j\ITomega\MATu{K}{\ITvarphi}]^{-1}
    \MAT{X}]^{-1}
    }
 
\newcommand{\EQdefAHimpedanceKphi}  
    {
    \MATu{K}{\ITvarphi} &=
    \MATu{M}{\ITmu}+\MATu{Q}{\TRANS}[
    \MATu{K}{\ITnu}+j\ITomega\MATu{M}{\ITsigma}
    ]^{-1}\MAT{Q}
    } 
 
\newcommand{\EqdefAHimpedanceTaylorZ}  
    {
    \MAT{Z}({j\ITomega}) \approx
    \MAT{Z}(0)
    + {j\ITomega}\left.\frac{\partial\MAT{Z}({j\ITomega})}{\partial\SCA{j\ITomega}}
    \right\rvert_{\CORRI{j}\SCA{\ITomega}=0}   
    }

\newcommand{\EqdefAHrelationUsIs}  
    {
    \VECl{u}{\RM{s}}(t) &\approx
    \MAT{R}\VECl{i}{\RM{s}}(t)+
    \MAT{L}\dx{t}\VECl{i}{\RM{s}}(t)
    }

\newcommand{\EqdefAHresistanceR}  
    {
    \MAT{R} &=
    [\MATu{X}{\TRANS}
    [\MATu{K}{\ITrho}]^{-1}\MAT{X}
    ]^{-1}
    }

\newcommand{\EqdefAHinductanceL}  
    {
    \MAT{L} &=
    \MAT{R}\MATu{X}{\TRANS}[
    \MATu{K}{\ITrho}]^{-1}
    [\MATu{M}{\ITmu}+\MATu{Q}{\TRANS}[\MATu{K}{\ITnu}]^{-1}\MAT{Q}]
    [\MATu{K}{\ITrho}]^{-1}\MAT{X}\MAT{R}
    }


\newcommand{\EQconstraintThinShellIsourcePort}
  {    
  \CORRII{\int\limits_{\SCAlu{\RMGamma}{\RM{H}}{r}}
  \VECu{\ITchi}{r}\cdot\VECu{K}{r}
  \D{\SCA{\RMGamma}}} 
  }  
      

\newcommand{\EQdefHThinShelldiscrete}
    {
    \VECu{H}{r} =
    \sum_{q}\VEClu{v}{q}{r}\SCAlu{h}{q}{r},\ \ 
    \VEClu{v}{q}{r} =
    \frac{\SCAlu{N}{q}{r}\left(\SCAlu{l}{\CORRII{\RMGamma}}{r},\SCAlu{m}{\CORRII{\RMGamma}}{r}\right)}{\SCAlu{\RMdelta}{\RM{t}}{r}}
    \VEClu{n}{\CORRII{\RMGamma}}{r}
    }

\newcommand{\EQmatAHNORMTdiscrete}
    { 
    \newcommand{\KA}{\MATu{K}{\ITnu}\!\!+\!\MATu{M}{\ITsigma}\dx{t}}
    \newcommand{\Q}{-\MAT{\BAR{Q}}}
    \newcommand{\QT}{\MATu{\BAR{Q}}{\TRANS}\dx{t}}   
    \newcommand{\M}{\MATu{\BAR{K}}{\ITrho}}
    \newcommand{\X}{-\MAT{\BAR{X}}}
    \newcommand{\XT}{\MATu{\BAR{X}}{\TRANS}}
    \newcommand{\NULL}{\MAT{0}}   
    \newcommand{\T}{\MATu{K}{\ITkappa}\!\!+\!\MATu{M}{\ITrho}\dx{t}}   
    \renewcommand*{\arraystretch}{1.25}    
    \begin{bmatrix} 
        \!\KA   & \!\!\!\!\Q    & \!\!\!\!\NULL & \!\!\!\!\NULL \!\\ 
        \!\QT   & \!\!\!\!\M    & \!\!\!\!\X    & \!\!\!\!\NULL \!\\ 
        \!\NULL & \!\!\!\!\XT   & \!\!\!\!\NULL & \!\!\!\!\NULL \!\\ 
        \!\NULL & \!\!\!\!\NULL & \!\!\!\!\NULL & \!\!\!\!\T    \!\\[3pt]     
    \end{bmatrix} 
    \!\!\cdot\!\!
    \begin{bmatrix}  
         \VEC{a}          \\
         \VEC{h}          \\
         \VECl{u}{\RM{s}} \\
         \VEC{t}          \\[3pt]
    \end{bmatrix}  
    \!\!=\!\!
    \begin{bmatrix}   
         \NULL                \! \\
         \NULL                \! \\
         \VECl{i}{\RM{s}}     \!  \\
         \VEC{\BAR{q}}(\cdot) \! \\[3pt]
    \end{bmatrix}
    }
 
\newcommand{\EQdiscreteCoeffAHNORMlabelQ}
    {
    (\SCA{\BAR{Q}})_{i,q}^{r} &=
    \CORRI{\SCAlu{\RMdelta}{\RM{t}}{r}}
    \int\limits_{\SCAlu{\RMGamma}{\RM{H}}{r}} 
    (\CURL\VEClu{v}{q}{r})\cdot\VECl{w}{i}
    \D{\SCA{\RMGamma}}
    }

\newcommand{\EQdiscreteCoeffAHNORMlabelKrho}
    {
    (\SCAu{\BAR{K}}{\ITrho})_{p,q}^{r} &=
    \CORRII{({\SCAlu{\RMdelta}{\RM{t}}{r}})^2}
    \int\limits_{\SCAlu{\RMGamma}{\RM{H}}{r}} 
    (\SCAlu{\ITrho}{\CORRII{\RMGamma}_{\RM{eq}}}{r}\CURL\VEClu{v}{q}{r})
    \cdot\CURL\VEClu{v}{p}{r}
    \D{\SCA{\RMGamma}}
    }
    
\newcommand{\EQdiscreteCoeffAHNORMlabelCHIOLD}
    {
    (\SCA{\BAR{X}})_{p}^{r} &=
    \CORRI{\SCAlu{\RMdelta}{\RM{t}}{r}}
    \int\limits_{\SCAlu{\ell}{\RM{J}}{r}}
    [(\CURL\VEClu{v}{p}{r})\CORRI{\times\VEClu{n}{\CORRII{\RMGamma}}{r}}]
    \cdot\D{\VEC{{\ell}}}
    } 
    
\newcommand{\EQdiscreteCoeffAHNORMlabelCHI}
    {
    (\SCA{\BAR{X}})_{p}^{r} &=
    \CORRI{\SCAlu{\RMdelta}{\RM{t}}{r}}
    \CORRII{\int\limits_{\SCAlu{\RMGamma}{\RM{H}}{r}}
    \VEClu{x}{p}{r}\cdot(\CURL\VEClu{v}{p}{r})
    \D{\SCA{\RMGamma}}}
    } 
    
 \newcommand{\EQdiscreteCoeffAHNORMlabelPJ}
    {
    (\SCA{\BAR{q}}(\cdot))_{m} &=
    \CORRI{\SCAlu{\RMdelta}{\RM{t}}{r}}
    \int\limits_{\SCAlu{\RMGamma}{\RM{H}}{r}}\!\! 
    \SCA{q}(\cdot)\SCAl{N}{m}
    \D{\SCA{\RMGamma}}
    }
       

\newcommand{\eqCurrentSharingKVL}
    {
    \SCAlu{\ITrho}{\RMGamma_{\RM{s}}}{r}\SCAu{\ITlambda}{r}\VECu{K}{r}=
    \SCAlu{\ITrho}{\RMGamma_{\RM{c}}}{r}(1-\SCAu{\ITlambda}{r})\VECu{K}{r}
    }

\newcommand{\eqCurrentSharingRhoScSurf}    
    {
    \SCAlu{\ITrho}{\RMGamma_{\RM{s}}}{r}(\SCAu{\ITlambda}{r}) &= 
    \frac{\SCAl{E}{\RM{crit}}}{\SCAlu{K}{\RM{crit}}{r}}
    \left(\SCAu{\ITlambda}{r}\SCAu{\RM{k}}{r}\right)^{\SCA{n}-1}
    }
    
\newcommand{\EQdefCriticalCurrentIndex} 
    {
    \SCAu{k}{r}=
    \frac{{|\VECu{K}{r}|}}{{\SCAlu{K}{\RM{crit}}{r}}}
    }
    
\newcommand{\eqCurrentSharingRhoNcSurf}    
    {
    \SCAlu{\ITrho}{\Gamma_{\RM{c}}}{r} &=
    \left(\sum_{i}^{\RM{N_c}}
    \frac{\CORRII{\SCAlu{\RMdelta}{\RM{t_c}}{r}}}{\SCAlu{\ITrho}{\RM{c},i}{r}}\right)^{-1}
    } 
 
 \newcommand{\eqCurrentSharingLambda}    
    {
    f(\SCAu{\ITlambda}{r})=\SCAu{\ITalpha}{r}(\SCAu{\ITlambda}{r})^{\SCA{n}}+\SCAu{\ITlambda}{r}-1
    } 

 \newcommand{\eqCurrentSharingAlpha}    
    {
    \SCAu{\ITalpha}{r}=
    \left.\frac{\SCAlu{\ITrho}{\RMGamma_{\RM{s}}}{r}(\SCAu{\ITlambda}{r})}{\SCAlu{\ITrho}{\RMGamma_{\RM{c}}}{r}}\right\rvert_{\SCAu{\ITlambda}{r}=1}
    } 

\newcommand{\eqCurrentSharingRhoEq}      
    {        
    \SCAlu{\ITrho}{\CORRII{\RMGamma}_{\RM{eq}}}{r}=
    \left(\frac{1}{\SCAlu{\ITrho}{\RMGamma_{\RM{s}}}{r}}+\frac{1}{\SCAlu{\ITrho}{\RMGamma_{\RM{c}}}{r}}\right)^{-1}. 
    }


\section{Introduction} 
    \label{SEC_Introduction}

\begin{figure}[tb]
  \centering
	\includegraphics[width=8.0cm]{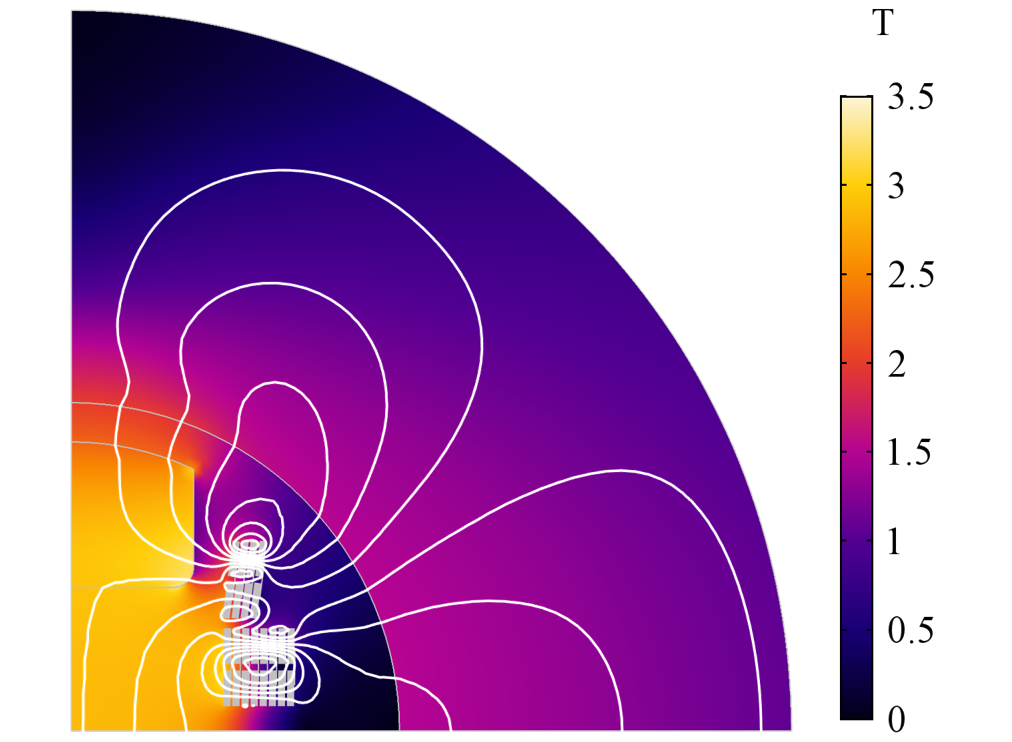}
	\caption{2D cross section \CORRI{showing one quadrant} of the \FEATHER magnet. \CORRI{The colormap illustrates} the magnetic flux density in T, at $\SI{6}{\kilo\ampere}$, during a linear ramp-up of \CORRII{$\SI{50}{\ampere\per\second}$}. The contour plot gives the magnetic flux lines caused by the eddy currents in the coil.}
	\label{FIG01_FeatherM2MagneticField}
\end{figure}

Particle colliders for high-energy physics are powerful tools for investigating the fundamental structure of matter. Circular accelerators such as the Large Hadron Collider ({LHC})~\cite{bruning2004lhc} rely on high-field magnets based on low-temperature superconductors ({LTS}), which  exhibit field-dependent superconducting properties~\cite{wilson1983superconducting}. This imposes a practical limit in the achievable magnetic field in the magnet aperture of about $\SI{8}{\tesla}$ for $\RM{Nb\CORRII{{\HYPHEN}}Ti}$, and $\SI{16}{\tesla}$ for $\RM{Nb_{3}Sn}$. In order to overcome these limits, the {EuCARD-2}~\cite{rossi2015eucard} and {ARIES}~\cite{ARIES2019website} projects aim for a technology shift, introducing high-temperature superconducting (HTS) tapes in accelerator magnets. Numerical methods play a key role in the development of HTS-based applications~\cite{grilli2013computation}, being crucial for analyzing the complex magneto-thermal dynamics occurring within HTS high-field magnets (e.g., the \FEATHER magnet~\CORRII{\cite{van2014study}},~\cite{van2018powering} Fig.~\ref{FIG01_FeatherM2MagneticField}). 

\CORRI{Among the available formulations for HTS materials, the so-called coupled $\VEC{A}\HYPHEN\VEC{H}$ formulation~\cite{dular1997magnetostatic,biro1999edge} is of relevance. The formulation solves time-domain eddy current problems by means of a domain decomposition. The field problem is solved for} the magnetic field strength $\VEC{H}$ in the regions containing the superconducting materials, and for the magnetic vector potential $\VEC{A}$ \CORRI{elsewhere}. For this particular application range, the \CORRI{increased complexity due to the domain decomposition is compensated by several advantages with respect to the canonical methods. In detail, one can avoid} the numerical instability of the \CORRI{monolithic} $\VEC{A}$-$\SCA{\ITvarphi}$ formulation related to the vanishing resistivity of the superconductor\CORRI{~\cite{ruiz2004numerical}}, the increased computational cost of the \CORRI{monolithic} $\VEC{H}$ formulation related to the introduction of unnecessary degrees of freedom~\cite{lahtinen2015finite}, and the \CORRI{increased complexity of} the \CORRI{coupled} $\VEC{T}$-\CORRII{$\SCA{\RMOmega}$} formulation related to the \CORRI{introduction} of the cohomology basis \CORRI{functions}~\cite{lahtinen2015finite}. \CORRI{The $\VEC{A}\HYPHEN\VEC{H}$ formulation has been recently applied to the simulation of both, rotating electrical machines containing superconducting windings~\cite{brambilla2018finite}, and superconducting materials~\cite{dular2019}.}

\CORRI{The objective of this work is to extend the capabilities of the coupled $\VEC{A}\HYPHEN\VEC{H}$ formulation~\cite{dular1997magnetostatic,biro1999edge}, providing an implementation in the finite element method (FEM) with mixed elements~\cite{bossavit1988rationale,dular1994mixed}, and showing that the formulation can be used for the simulation of HTS magnets.} In detail, we derive a general 3D time-domain eddy current formulation in solid superconductors with arbitrary excitation. \CORRI{Then}, a field-circuit coupling interface suitable for the co-simulation of superconducting magnets~\cite{bortot2018steam} with a waveform relaxation scheme~\cite{schops2010cosimulation} is proposed. \CORRI{Furthermore, the thin shell approximation for eddy currents similar to~\CORRI{\cite{carpenter1977comparison,rodger1988finite,krahenbuhl1993thin}} is derived from the general formulation, and it is applied to the source regions consisting of superconducting tapes with high aspect ratio. As the thickness of the tapes is neglected, this allows to solve the current sharing regime~\cite{wilson1983superconducting} within the tapes in a computationally efficient way.} \CORRI{The consistency of the formulation is verified and validated against available reference models. Finally the formulation is upscaled to a 2D model of the \FEATHER magnet~\cite{van2018powering} for which the time-domain electrodynamic phenomena occurring in the superconducting coil are quantified.}


\section{Method}
    \label{SEC_Method}
    
\begin{figure}[tb]
  \centering
	\includegraphics[width=8.5cm]{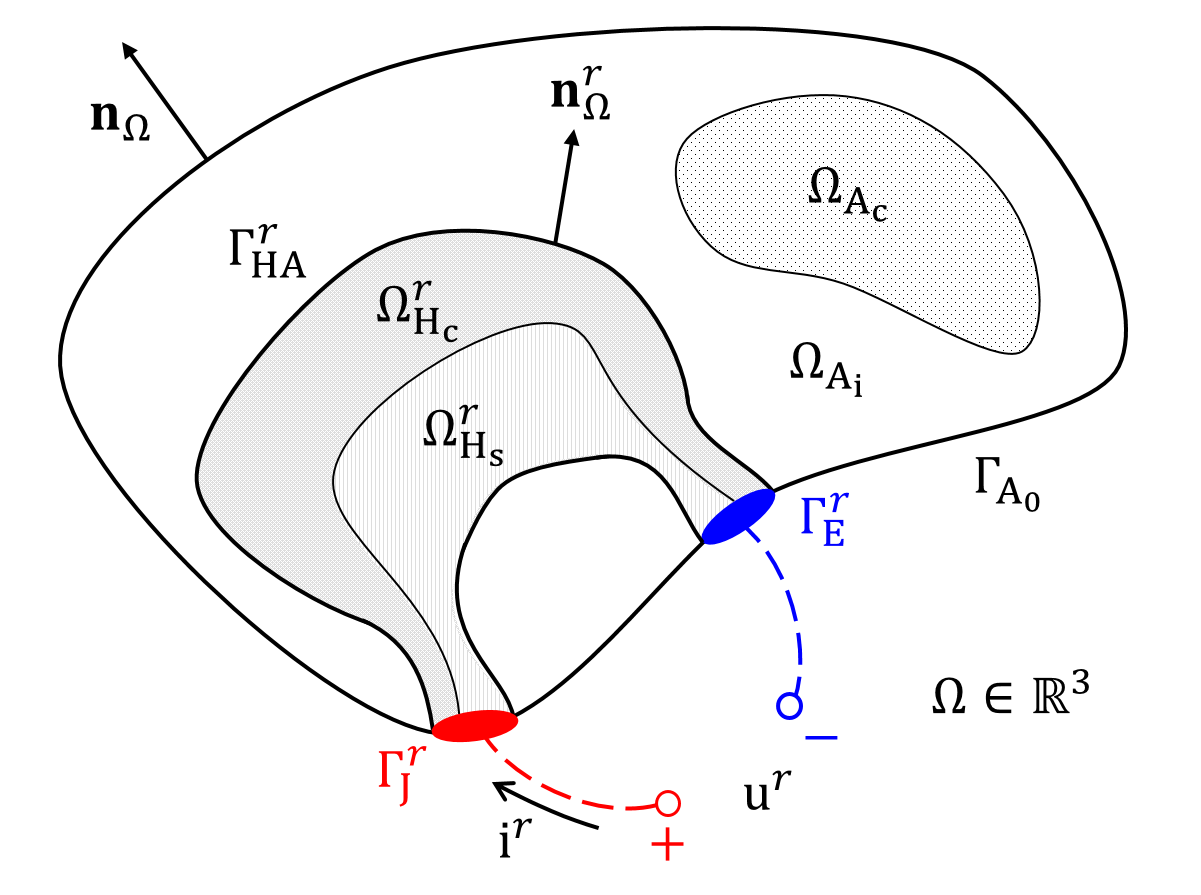}
	\caption{Decomposition of the domain $\SCA{\RMOmega}$ into $\SCAlu{\RMOmega}{\RM{H}}{r}$ and $\SCAl{\RMOmega}{\RM{A}}$, representing the source and source-free domains, respectively. The domain $\SCAlu{\RMOmega}{\RM{H}}{r}$ is bounded by $\SCAlu{\BAR{\RMGamma}}{\RM{H}}{r}=\SCAlu{\BAR{\RMGamma}}{\RM{HA}}{r}\!\cup\SCAlu{\BAR{\RMGamma}}{\RM{E}}{r}\cup\SCAlu{\BAR{\RMGamma}}{\RM{J}}{r}$, where \CORRII{$\SCAlu{\RMGamma}{\RM{HA}}{r}$} denotes the interface between the domains \CORRII{$\SCAlu{\RMOmega}{\RM{H}}{r}$ and $\SCAl{\RMOmega}{\RM{A}}$}, and $\SCAlu{\RMGamma}{\RM{E}}{r}$ and $\SCAlu{\RMGamma}{\RM{J}}{r}$ represent the electric ports. The domain $\SCAl{\RMOmega}{\RM{A}}$ is bounded by $\SCAl{\BAR{\RMGamma}}{\RM{A}}=\CORRII{\SCAl{\BAR{\RMGamma}}{\RM{A_0}}}\cup\CORRII{\SCAlu{\BAR{\RMGamma}}{\RM{HA}}{r}}$. 
	The dashed lines represent the \CORRII{electrical connections}, which may be used to impose a source term on $\SCAlu{\RMOmega}{\RM{H}}{r}$, either as a voltage $\SCAu{u}{r}$ or a current $\SCAu{i}{r}$.
    }
	\label{FIG02_DomainDecomposition}
\end{figure}

Superconducting materials based on HTS technology exhibit a strongly nonlinear electrical behavior. \CORRI{Their resistivity can be represented by a \CORRI{phenomenological percolation-depinning law~\cite{yamafuji1997current}}, which is a generalization of the power law~\cite{rhyner1993magnetic}. For practical applications, the two laws are in agreement~\cite{sirois2018comparison}, and the power-law behavior is considered.} The law shows a dependency on both, the current density $\VEC{J}$, and the critical field strength $\SCAl{E}{\RM{crit}}$. The most common model for the specific resistivity $\SCA{\ITrho}$ for superconducting materials\CORRI{~\cite{kim1965flux}} reads
    \begin{align}
        & \EQdefJpowerLaw,
        \label{EQdefJpowerLaw}
    \end{align}
where the critical current density $\SCAl{J}{\RM{crit}}$ and the power-law exponent $\SCA{n}$ are material-dependent parameters, and $\SCAl{E}{\RM{crit}}$ is chosen as \CORRII{$\SCAl{E}{\RM{crit}}=\SI{1e-4}{\volt\per\meter}$}~\cite{dew1988model}. 

The power law in~\eqref{EQdefJpowerLaw} prescribes a vanishing resistivity $\SCA{\ITrho}\to0$ for $|\VEC{J}|\ll\SCAl{J}{\RM{crit}}$. On the one hand, the field formulation should not use the electrical conductivity $\SCA{\ITsigma}=\SCAu{\ITrho}{-1}$ in the superconducting domains, since $\SCA{\ITsigma}\to\infty$. On the other hand, if non-conducting regions are present, $\SCA{\ITrho}$ should not be used either, since $\SCA{\ITrho}\to\infty$. This impasse is overcome by using a domain decomposition strategy.


\subsection{Domain Decomposition}
    \label{SUBSEC_DomainDecomposition}

The domain decomposition strategy is based on the topological separation of the field source domain from the source-free one, and it is implemented by following the structure of a superconducting accelerator magnet. The source domain represents the excitation coil, \CORRII{which provides} the magnetic field source. The coil is composed by $\SCAl{N}{\RM{r}}$ electrically independent windings. The source-free domain represents the rest of the magnet consisting of the iron yoke, the mechanical supports and the air \CORRI{regions}. \CORRII{The electrical insulation of the excitation coil ensures the electrical separation between the source and source-free domains. In particular, magnetic materials and source domains do not touch}. The domain decomposition is formalized as follows. 


The domain 
$\SCA{\RMOmega}\subset\mathbb{R}^3$ bounded by $\SCA{\BAR{\RMGamma}}=\dxP{}\SCA{\RMOmega}$ is decomposed into the source domain $\SCAl{\RMOmega}{\RM{H}}$ \CORRII{bounded by $\SCAl{\BAR{\RMGamma}}{\RM{H}}$}, and the source-free domain $\SCAl{\RMOmega}{\RM{A}}$ \CORRII{bounded by $\SCAl{\BAR{\RMGamma}}{\RM{A}}$}. The domain decomposition is such that $\SCAl{\BAR{\RMOmega}}{\RM{H}}\cup\SCAl{\BAR{\RMOmega}}{\RM{A}}=\CORRII{\SCA{\BAR{\RMOmega}}}$. The source domain is further subdivided into $\SCAl{N}{\RM{r}}$ non-intersecting domains $\SCAlu{\RMOmega}{\RM{H}}{r}$ representing the coil windings, as $\SCAl{\BAR{\RMOmega}}{\RM{H}}=\bigcup_{r=1}^{\SCAl{N}{\RM{r}}}\SCAlu{\BAR{\RMOmega}}{\RM{H}}{r}$ \CORRII{bounded by} $\SCAl{\BAR{\RMGamma}}{\RM{H}}=\bigcup_{r=1}^{\SCAl{N}{\RM{r}}}\SCAlu{\BAR{\RMGamma }}{\RM{H}}{r}$.

The domain decomposition strategy is detailed in Fig.~\ref{FIG02_DomainDecomposition}: for the sake of clarity, only the \RTHINDEX source domain is represented. 
Each source domain $\SCAlu{\RMOmega}{\RM{H}}{r}$ is oriented with the unit pointing vector \CORRII{$\VEClu{n}{\RMOmega}{r}$} and it is contoured by 
$\SCAlu{\BAR{\RMGamma}}{\RM{H}}{r}=\SCAlu{\BAR{\RMGamma}}{\RM{HA}}{r}\cup\SCAlu{\BAR{\RMGamma}}{\RM{E}}{r}\cup\SCAlu{\BAR{\RMGamma}}{\RM{J}}{r}$, 
where \CORRII{$\SCAlu{\RMGamma}{\RM{HA}}{r}$} is the interface boundary with $\SCAl{\RMOmega}{\RM{A}}$, and $\SCAlu{\RMGamma}{\RM{E}}{r}$ and $\SCAlu{\RMGamma}{\RM{J}}{r}$ \CORRI{are} the electrical ports provided to each winding. The source-free domain $\SCAl{\RMOmega}{\RM{A}}$ is oriented with the outward pointing vector $\VECl{n}{\RMOmega}$, and it is contoured by
$\SCAl{\BAR{\RMGamma}}{\RM{A}}=\CORRII{\SCAl{\BAR{\RMGamma}}{\RM{A_0}}}\cup\CORRII{\SCAl{\BAR{\RMGamma}}{\RM{HA}}}$,
\CORRII{where $\SCAl{\BAR{\RMGamma}}{\RM{A_0}}$ is the boundary of $\SCA{\RMOmega}$ without the electrical ports, and 
$\SCAl{\BAR{\RMGamma}}{\RM{HA}}=\bigcup_{r=1}^{\SCAl{N}{\RM{r}}}\SCAlu{\BAR{\RMGamma}}{\RM{HA}}{r}$} is the cumulative interface. Each source domain may contain both, superconducting and normal conducting sub-domains, represented by $\SCAlu{\RMOmega}{\RM{H_s}}{r}$ and $\SCAlu{\RMOmega}{\RM{H_c}}{r}$, respectively. The domain $\SCAl{\RMOmega}{\RM{A}}$ may contain both, normal conducting and insulating sub-domains, represented by $\SCAl{\RMOmega}{\RM{A_c}}$ and $\SCAl{\RMOmega}{\RM{A_i}}$, respectively.

\subsection{Full 3D Formulation}
    \label{SUBSEC_Full3DFormulation}

The formulation is defined in the computational domain $\SCA{\RMOmega}$ under magnetoquasistatic assumptions, solving for the magnetic vector potential $\VEC{A}$ in $\SCAl{\RMOmega}{\RM{A}}$ and for the magnetic field strength $\VEC{H}$ in  $\SCAl{\RMOmega}{\RM{H}}$. To avoid ambiguity, the field variables are restricted as $\VEC{A}=\VEC{A}|_{\SCAl{\RMOmega}{\RM{A}}}$ and $\VEC{H}=\VEC{H}|_{\SCAl{\RMOmega}{\RM{H}}}$. In the following, all the field quantities are assumed to be space- and time-dependent. The gauge freedom is exploited such that $\VEC{A}$ is replaced by the reduced magnetic vector potential $\VECu{A}{\star}$~\cite{emson1983optimal}. 
This choice gives a unique solution for the normal conducting domain $\SCAl{\RMOmega}{\RM{A_c}}$, whereas an extra gauge condition is required for the nonconducting domain $\SCAl{\RMOmega}{\RM{A_i}}$. The magnetic permeability $\SCA{\ITmu}$ determining the magnetic constitutive law is assumed constant in the domains $\SCAl{\RMOmega}{\RM{H}}$ and in $\SCAl{\RMOmega}{\RM{A_i}}$, \CORRII{whereas} a nonlinear field dependency is assumed in $\SCAl{\RMOmega}{\RM{A_c}}$, i.e., $\SCA{\ITmu}(\VEC{B})$ where $\VEC{B}$ is the magnetic flux density. 

As the source domain is composed by $\SCAl{N}{\RM{r}}$ independent windings, the magnetic field strength is given as $\VEC{H}=\sum_{r}^{\SCAl{N}{\RM{r}}}\VECu{H}{r}$. Moreover, each winding $\SCAlu{\RMOmega}{\RM{H}}{r}$ may carry independent source voltages and currents, i.e., $\EQdefUs$ and $\EQdefIs$, respectively. To achieve this, the windings are modeled by dedicated voltage distribution functions for solid conductors $\EQdefCHI$, \CORRII{as in}~\cite{schops2013winding}. Each function is defined as $\VECu{\ITchi}{r}=-\GRAD\SCAu{\ITxi}{r}$ where $\SCAu{\ITxi}{r}$ is the electric scalar potential. \CORRI{Following~\cite{rodriguez2008voltage}, the potential is obtained \CORRII{separately}, solving a unitary current-flow problem for each \RTHINDEX winding composing the coil}. With this definition, $\VECu{\ITchi}{r}$ can be interpreted as a per-unit electric field strength. Next, the electric field strength $\VEC{E}$ is split into $\EQdefE$, where the source \CORRII{field}
        \begin{align}
        \EQdefEs \label{EQdefEs}
        \end{align}
\CORRI{represents the contribution of the external source voltage $\VECl{u}{\RM{s}}$ which is applied to the windings}. Thus, the overall electric field strength distributed among the $\SCAl{N}{\RM{r}}$ windings is $\VEC{E}=\sum\nolimits_{r}^{\SCAl{N}{\RM{r}}}\VECu{E}{r}$, where $\EQdefEr$.

Alternatively, if the source current $\VECl{i}{\RM{s}}$ is prescribed, the source voltage becomes an algebraic unknown, and one constraint equation is added (i.e., serves as a Lagrange multiplier) for each of the independent currents in $\VECl{i}{\RM{s}}$. With the previous observations, the strong formulation of the field problem reads: find $\VECu{A}{\star}$, $\VECu{H}{r}$ and $\SCAu{u}{r}$, for $r=1,...,\SCAl{N}{\RM{r}}$, such that 
    \begin{align}
            \EQstrongAampereMaxwell\ \text{in}\ \SCAl{\RMOmega}{\RM{A}}, 
                \label{EQstrongAampereMaxwell}\\
            \EQstrongHfaraday\ \text{in}\ \SCAlu{\RMOmega}{\RM{H}}{r}, 
                \label{EQstrongHfaraday}\\
            \EQconstraintIsource &= \SCAu{i}{r}, 
                \label{EQconstraintIsource}
    \end{align}
with the Dirichlet boundary conditions
    \begin{align}
            \EQdirichletAHtangA\ \text{on}\ \CORRII{\SCAl{\RMGamma}{\RM{A_0}}},                    
                \label{EQdirichletAHtangA}\\
            \EQdirichletAHtangE\ \text{on}\ \SCAlu{\RMGamma}{\RM{E}}{r},\ \SCAlu{\RMGamma}{\RM{J}}{r}\CORRII{.}
                \label{EQdirichletAHtangE}
    \end{align}

To ensure the consistency of the overall solution, the fields $\VECu{A}{\star}$ and $\VEC{H}$ are linked by appropriate interface conditions at each boundary $\SCAlu{\RMGamma}{\RM{HA}}{r}$. In detail, the continuity of the normal components of $\VEC{J}$ and $\VEC{B}$, and the tangential components of $\VEC{H}$ and $\VEC{E}$ must be ensured. The interface conditions are given on $\SCAlu{\RMGamma}{\RM{HA}}{r}$ for $r=1,...,\SCAl{N}{\RM{r}}$, as
        \begin{align}
        \EQinterfaceAHnormJ, 
            \label{EQinterfaceAHnormJ} \\
        \EQinterfaceAHnormB, 
            \label{EQinterfaceAHnormB} \\
        \EQinterfaceAHtangH, 
            \label{EQinterfaceAHtangH} \\
        \EQinterfaceAHtangE. 
            \label{EQinterfaceAHtangE}
        \end{align}

The materials typically used in superconducting devices show magnetic- and temperature-dependent physical properties. This requires to solve the temperature field $\SCA{T}$ together with the magnetoquasistatic problem described by~\eqref{EQstrongAampereMaxwell}$-$\eqref{EQconstraintIsource}. In particular, the temperature influences both, the critical current density of the superconducting materials, and the resistivity of the normal conducting materials. 
This, in turn, determines the Joule losses $\EQdefJjouleLoss$ occurring in the conducting domains, \CORRII{and acting as the main heat source term}.
It is worth noting that the coupling current \CORRII{phenomena}~\cite{wilson1983superconducting} \CORRII{occurring} in the source domain \CORRII{provide} a second-order \CORRII{loss contribution} with respect to the \CORRII{one from} the conduction current~\cite{van2016high}. 
The temperature is obtained by solving the heat balance equation
    \begin{align}
        \EQstrongTheatBalance\ \text{in}\ \SCA{\RMOmega}
        \label{EQstrongTheatBalance}
    \end{align}
where $\SCAl{\ITrho}{\RM{m}}$ is the mass density, $\SCAl{C}{\RM{p}}$ the heat capacity, and $\SCA{\ITkappa}$ the thermal conductivity. The Neumann boundary condition $\SCA{\ITkappa}\GRAD\SCA{T}\cdot\VECl{n}{\RMOmega}=0$ is applied on $\SCA{\RMGamma}$. \CORRII{If} the source domain is approximated as adiabatic, the Neumann boundary condition is moved to $\SCAl{\RMGamma}{\RM{H}}$, thereby reducing the model complexity.


\subsection{Weak Formulation}
    \label{SUBSEC_WeakFormulation}

The weak formulation of the field problem is obtained by applying the Ritz-Galerkin method (e.g.,~\cite{bossavit1998}). The equations~\eqref{EQstrongAampereMaxwell} and~\eqref{EQstrongHfaraday} are weighted with the vector functions $\VECl{w}{i}$ and $\VECl{v}{p}$, respectively, \CORRII{whereas} the equation~\eqref{EQstrongTheatBalance} is weighted with the scalar functions $\SCAl{N}{m}$. Integrating the weighted equations on $\SCA{\RMOmega}$ leads to the following problem: find $\VECu{A}{\star}$, $\VECu{H}{r}$ and $\SCAu{u}{r}$, \CORRII{for $r=1,...,\SCAl{N}{\RM{r}}$,} such that
    \begin{align}
        \begin{split}
            \EQweakAampereMaxwell, 
                \label{EQweakAampereMaxwell}
        \end{split}
        \\[1ex]
        \begin{split}
            \EQweakHfaraday, 
                \label{EQweakHfaraday}
        \end{split}
        \\[1ex]
        \begin{split}
            \EQweakTheatBalance, 
                \label{EQweakTheatBalance}
        \end{split}
    \end{align}
for all test functions $\VECl{w}{i}$, $\VEClu{v}{p}{r}$, $\SCAl{N}{m}$. The Dirichlet boundary conditions~\eqref{EQdirichletAHtangA} and~\eqref{EQdirichletAHtangE} are assumed to be incorporated into the function spaces. The third integral in~\eqref{EQweakAampereMaxwell} and the fourth integral in~\eqref{EQweakHfaraday} provide the natural coupling interface exploited by the coupled field formulation. It is worth noting that following~\cite{rodriguez2008voltage}, the constraint condition for the current in~\eqref{EQconstraintIsource} can be reformulated as
    \begin{align}
            \EQconstraintIsource=\EQconstraintIsourcePort,
            \label{EQconstraintIsourcePort}
    \end{align}
limiting the support of the integral to $\SCAlu{\RMGamma}{\RM{J}}{r}$, thus reducing the complexity of the constraint.


\subsection{Discretization}
    \label{SUBSEC_Discretization}

The fields $\VECu{A}{\star}$ and $\VECu{H}{r}$ are approximated for $r=1,...,\SCAl{N}{\RM{r}}$ by a finite set of N\'{e}d\'{e}lec-type shape functions $\VECl{w}{j}$ and $\VEClu{v}{q}{r}$ as $\EQdefAdiscrete$ and $\EQdefHdiscrete$. \CORRII{The voltage distribution function is known and it is discretized by nodal shape functions $\SCAl{N}{p}$, as $\EQdefCHIdiscrete$}. The field $\SCA{T}$ is discretized by nodal shape functions $\SCAl{N}{\CORRII{n}}$, as $\EQdefTdiscrete$. The unknowns $\SCAl{a}{j}$, $\SCAlu{h}{q}{r}$ and $\SCAl{t}{n}$ are the degrees of freedom for $\VECu{A}{\star}$, $\VECu{H}{r}$ and $\SCA{T}$, respectively. As a consequence, the unknown field $\VEC{h}$ is given as $\VECu{h}{\TRANS}=[(\VECu{h}{1})^{\TRANS},\ldots,(\VECu{h}{\SCAl{N}{\RM{r}}})^{\TRANS}]$. 

The fields are replaced in~\eqref{EQweakAampereMaxwell}$-$\eqref{EQweakTheatBalance} with their finite dimensional counterparts. The interface conditions~\eqref{EQinterfaceAHtangH} and~\eqref{EQinterfaceAHtangE} are explicitly imposed for the tangential component of $\VECu{H}{r}$ and $\VECu{E}{r}$, respectively. The continuity of the normal component of the current density and  magnetic flux density, given respectively by~\eqref{EQinterfaceAHnormJ} and~\eqref{EQinterfaceAHnormB}, is satisfied by choosing suitable discretization functions based on Whitney edge elements~\cite{barton1987new}. The following discrete problem is obtained:
    \begin{align}
      \!\!\!\!\!\EQmatAHTdiscrete\!.\!\!\!
        \label{EQmatAHTdiscrete}
    \end{align}
In detail, $\MATu{K}{\star}$ and $\MATu{M}{\star}$ represent the discrete counterparts of differential operators and material matrices, $\MAT{X}$ is the discrete representation of $\VEC{\ITchi}$ and \CORRII{$\SCA{q}(\VEC{a},\VEC{h})$ is the nonlinear discrete operator representing the Joule loss contribution.}
The coefficients of the matrices of reluctance $\MATu{K}{\ITnu}$, conductance $\MATu{M}{\ITsigma}$, interface coupling $\MAT{Q}{}{}$, resistance $\MATu{K}{\ITrho}$, permeance $\MATu{M}{\ITmu}$ and voltage coupling $\MAT{X}$ are given for $r=1,...,\SCAl{N}{\RM{r}}$ as
    \begin{align}
        \EQdiscreteCoeffAHlabelKnu,    \label{EQdiscreteCoeffAHlabelMnu}  \\
        \EQdiscreteCoeffAHlabelMsigma, \label{EQdiscreteCoeffAHlabelMs}   \\
        \EQdiscreteCoeffAHlabelQ,      \label{EQdiscreteCoeffAHlabelQ}    \\
        \EQdiscreteCoeffAHlabelKrho,   \label{EQdiscreteCoeffAHlabelMrho} \\
        \EQdiscreteCoeffAHlabelMmu,    \label{EQdiscreteCoeffAHlabelMmu} \\
        \EQdiscreteCoeffAHlabelCHI.    \label{EQdiscreteCoeffAHlabelCHI}
    \end{align}
The matrix coefficients of heat capacitance $\MATu{M}{\ITrho}$ and heat diffusion $\MATu{K}{\ITkappa}$, and the Joule loss \CORRII{contribution $\SCA{q}(\cdot)$} are given as
    \begin{align}
        \EQdiscreteCoeffTlabelKk,   \label{EQdiscreteCoeffTlabelKk} \\
        \EQdiscreteCoeffTlabelMrho, \label{EQdiscreteCoeffTlabelMrho} \\
        \EQdiscreteCoeffTlabelPjoule, \label{EQdiscreteCoeffTlabelPjoule}
    \end{align}
If initial conditions and currents $\VECl{i}{\RM{s}}$ are given, then the semi-discrete system~\eqref{EQmatAHTdiscrete} is ready to be solved by a time-stepping algorithm, e.g., the backward differentiation formula ({BDF})~\cite{ascher1998computer}. However, if the currents depend on a surrounding circuitry, then further derivations are necessary\CORRI{~\cite{bortot2018steam}}.


\subsection{Field-Circuit Coupling}
    \label{SUBSEC_FieldCircuitCoupling}
    
Each source domain $\SCAlu{\RMOmega}{\RM{H}}{r}$ is equipped with two electrical ports $\SCAlu{\RMGamma}{\RM{E}}{r}$ and $\SCAlu{\RMGamma}{\RM{J}}{r}$, providing the source terms to the field model. If an external electrical network is present, these ports can be exploited \CORRII{to apply} the source terms \CORRII{determined by} the network. \CORRII{For this reason}, a field-circuit coupling interface is introduced, allowing to connect the electric field strength and current density in each source domain to voltages and currents of an external circuit model. Among the possible coupling schemes, the co-simulation of the field and the circuit models using the waveform relaxation scheme~\cite{schops2011multiscale} is suitable for the magneto-thermal transients in high-field accelerator magnets~\cite{bortot2016consistent}. 

The field-circuit coupling interface is derived as Schwarz transmission condition for linear systems~\cite{al2014optimization,garcia2017optimized} which is optimized for co-simulation schemes, resulting in a speed-up of the convergence rate of the co-simulation algorithm. The coupling interface can be \CORRII{formalized} as a voltage-current relation $\VECl{u}{\RM{s}}=\MAT{Z}\VECl{i}{\RM{s}}$ for a multi-port electrical device, where the impedance $\MAT{Z}$ is a $\SCAl{N}{\RM{r}}\times\SCAl{N}{\RM{r}}$ dimensional matrix. The expression for $\MAT{Z}$ is obtained in frequency domain with $\ITomega=2\pi f$. The discrete counterpart of the magnetic vector potential $\VEC{a}$ is assumed to be gauged in $\SCAl{\RMOmega}{\RM{A}}$ (e.g., via a tree-cotree gauge condition~\cite{manges1995generalized}), such that the matrix $\MATu{K}{\ITnu}+j\ITomega\MATu{M}{\ITsigma}$ is positive-definite, thus invertible. Moreover, $\MATu{K}{\ITrho}$ is positive semidefinite, $\MATu{M}{\ITmu}$ is positive definite and $\MAT{Q}$ and $\MAT{X}$ have full-column rank. Thus, it is possible to use the Schur complement in~\eqref{EQmatAHTdiscrete} to derive the voltage-current relation, leading to 
    \begin{align}
        \EQdefAHimpedanceZ,  
            \label{EQdefAHimpedanceZ}
    \end{align}
where the reluctance matrix $\MATu{K}{\ITvarphi}$ is defined as 
    \begin{align}
        \EQdefAHimpedanceKphi.  
            \label{EQdefAHimpedanceKphi}
    \end{align}
In frequency domain, \CORRII{the impedance in}~\eqref{EQdefAHimpedanceZ} is computable and provides the optimal transmission condition. However, when dealing with nonlinear systems in time-domain, the time derivatives contained in $\MAT{Z}$ need to be approximated. This is achieved by following~\cite{garcia2017optimized}, i.e., $\MAT{Z}$ is approximated by a Taylor expansion series truncated to the first order, as
    \begin{align}
        \EqdefAHimpedanceTaylorZ.
            \label{EqdefAHimpedanceTaylorZ}
    \end{align}
Using the result from ~\eqref{EQdefAHimpedanceZ} into~\eqref{EqdefAHimpedanceTaylorZ} leads to
    \begin{align}
        \EqdefAHrelationUsIs,
            \label{EqdefAHrelationUsIs}
    \end{align}
where $\MAT{R}$ and $\MAT{L}$ represent equivalent resistance and inductance matrices, respectively. The matrices are defined as
\allowdisplaybreaks[0]
    \begin{align}
        \!\!\EqdefAHresistanceR, 
            \label{EqdefAHresistanceR} \\
        \!\!\EqdefAHinductanceL, 
            \label{EqdefAHinductanceL} 
    \end{align}
\allowdisplaybreaks
and may be used to approximate the finite element model in the circuital counterpart, as an {RL}-series component. The low-order model introduced by~\eqref{EqdefAHimpedanceTaylorZ} disregards several effects, in particular the contribution  of the eddy currents $\MATu{M}{\ITsigma}$ occurring in the domains outside the source region. As the Taylor series is expanded around $\CORRI{j}\ITomega=0$, such approximation should be limited to low frequencies, i.e., small $\ITomega$.


\section{\CORRI{Thin Shell} Approximation}
    \label{SEC_ThinShellApproximation}

\begin{figure}[tb]
  \centering
	\includegraphics[width=8.0cm]{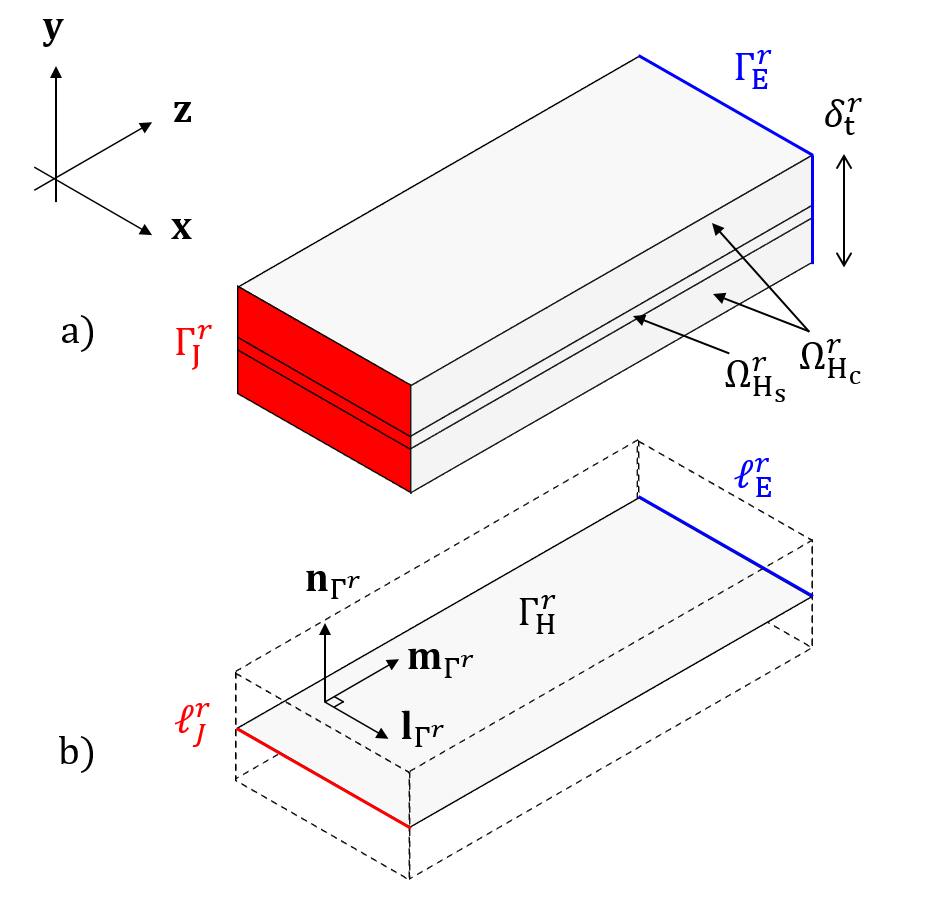}
	\caption{a) Example of a superconducting tape. The composite material structure is reflected by $\SCAlu{\BAR{\RMOmega}}{\RM{H}}{r}=\SCAlu{\BAR{\RMOmega}}{\RM{H_s}}{r}\cup\ \SCAlu{\BAR{\RMOmega}}{\RM{H_c}}{r}$, representing the superconducting and the \CORRII{normal} conducting layers, respectively. b) The thickness $\SCAlu{\RMdelta}{\RM{t}}{r}$ of the tape is neglected, leading to the collapse of the source domain into \CORRI{an equivalent thin shell} $\SCAlu{\RMGamma}{\RM{H}}{r}$ which accounts for the electrical behavior of both, $\SCAlu{\RMOmega}{\RM{H_s}}{r}$ and $\SCAlu{\RMOmega}{\RM{H_c}}{r}$. The electrical ports \CORRI{$\SCAlu{\RMGamma}{\RM{J}}{r}$ and $\SCAlu{\RMGamma}{\RM{E}}{r}$} are \CORRII{reduced to the edges} $\SCAlu{\ell}{\RM{E}}{r}$ and $\SCAlu{\ell}{\RM{J}}{r}$, respectively.}
	\label{FIG03_ThinShellApproximation}
    \vskip -0.25cm
\end{figure}

Superconducting tapes exhibit a layered structure of composite materials with high aspect ratio, up to two orders of magnitude. \CORRII{As example}, Fig.~\ref{FIG03_ThinShellApproximation}a represents the \CORRII{tape} $\SCAlu{\RMOmega}{\RM{H}}{r}$ \CORRII{of thickness $\SCAlu{\RMdelta}{\RM{t}}{r}$.} The \CORRII{tape} is composed by one superconducting layer $\SCAlu{\RMOmega}{\RM{H_s}}{r}$ \CORRII{of thickness $\SCAlu{\RMdelta}{\RM{t_s}}{r}$}, and a normal conducting layer $\SCAlu{\RMOmega}{\RM{H_c}}{r}$ \CORRII{of thickness $\SCAlu{\RMdelta}{\RM{t_c}}{r}$}, which is made of composite materials and provides mechanical and thermal support.

\CORRII{The high aspect ratio} of the superconducting tape justifies to approximate the domain as a thin shell, neglecting its thickness $\SCAlu{\RMdelta}{\RM{t}}{r}$. \CORRII{In this way, the tape can be represented by an equivalent surface  $\SCAlu{\RMGamma}{\RM{H}}{r}$ (see Fig.~\ref{FIG03_ThinShellApproximation}b)}. The current sharing regime between the conducting layers is resolved via the equivalent surface resistivity $\SCAlu{\ITrho}{\RMGamma_{\RM{eq}}}{r}$ (see Section~\ref{SUBSEC_CurrentSharing}), \CORRII{which homogenizes the resitivity the tape. At the same time,} the source domain $\SCAlu{\RMOmega}{\RM{H}}{r}$ \CORRII{is kept} as an insulating structural element (dashed line, Fig.~\ref{FIG03_ThinShellApproximation}b) accounted for in $\SCAl{\RMOmega}{\RM{A}}$. \CORRII{In this way,} the thermal contacts are preserved for the thermal problem in~\eqref{EQstrongTheatBalance}.
The implications \CORRI{on the coupled field formulation are discussed in the following.}


\subsection{Field Properties in the Shell}
    \label{SUBSEC_FieldPropertiesShell}
    
With respect to the global Cartesian coordinate system $(\VEC{x},\VEC{y},\VEC{z})$, a local coordinate system $(\VEClu{l}{\CORRII{\RMGamma}}{r},\VEClu{m}{\CORRII{\RMGamma}}{r},\VEClu{n}{\CORRII{\RMGamma}}{r})$ is introduced and oriented according to the tangential ($\VEClu{l}{\CORRII{\RMGamma}}{r}$ and $\VEClu{m}{\CORRII{\RMGamma}}{r}$) and normal ($\VEClu{n}{\CORRII{\RMGamma}}{r}$) directions of the \RTHINDEX tape. The local coordinate system is used to decompose the differential operators and vectors $\VEC{v}$ into their tangential (t) and normal (n) components, i.e., $\GRAD=\GRADTANG+\GRADNORM$ and $\VEC{v}=\VECl{v}{\RM{t}}+\VECl{v}{\RM{n}}$. \CORRI{The thin shell approximation is introduced by applying the following conditions. First, the current only flows along the shell, i.e.}
    \begin{align}
            \CORRI{\VEC{J}\cdot\VEClu{n}{\CORRII{\RMGamma}}{r}=0 \ \text{on}\ \SCAlu{\RMGamma}{\RM{H}}{r}.}  
                \label{EqThinShellConstaintJ}
    \end{align}
\CORRI{Second, the variation of all field quantities along the perpendicular direction of the tape $\VEClu{n}{\CORRII{\RMGamma}}{r}$ is zero, as}
    \begin{align}
            \CORRI{
            \GRAD\cdot=
            \left(
            \frac{\dxP{}\cdot}{\dxP{\SCAlu{l}{\CORRII{\RMGamma}}{r}}},
            \frac{\dxP{}\cdot}{\dxP{\SCAlu{m}{\CORRII{\RMGamma}}{r}}},
            0
            \right) 
            = \GRADTANG\cdot \ \text{on}\ \SCAlu{\RMGamma}{\RM{H}}{r}.
            }        
            \label{EqThinShellConstaintH}
    \end{align}
\CORRII{The conditions given by~\eqref{EqThinShellConstaintJ} and~\eqref{EqThinShellConstaintH} lead to define} the equivalent surface current density $\VECu{K}{r}$ in the shell as
    \begin{align}
            \CORRI{
            \VECu{K}{r}
            = \int\limits_{\SCAlu{\RMdelta}{\RM{t}}{r}}\VEC{J}\D{\SCA{\ell}}
            = \SCAlu{\RMdelta}{\RM{t}}{r}\CURLTANG\VEClu{H}{\RM{n}}{r}
            \ \text{on}\ \SCAlu{\RMGamma}{\RM{H}}{r},
            }
                \label{EQdefThinShellCurrentK}
    \end{align}
\CORRII{thus $\VECu{K}{r}$ is determined only by} the normal component $\VEClu{H}{\RM{n}}{r}$ of the magnetic field strength within the shell.


\subsection{Field Properties at the Shell Boundary}
    \label{SUBSEC_FieldPropertiesShellBoundary}
    
\CORRI{As the thin shell is a surface, special interface conditions \CORRII{are put in place} to \CORRII{ensure the consistency of the overall solution}. \CORRII{In detail}, the surface current density in \CORRII{\eqref{EQdefThinShellCurrentK}} introduces a discontinuity of the tangential magnetic field strength \CORRII{in $\SCAl{\RMOmega}{\RM{A}}$}, as
    \begin{align}
            \VECu{K}{r}=
            (\CORRII{\ITmu^{-1}}(\CURL\VEClu{A}{1}{\star}-\CURL\VEClu{A}{2}{\star})
            \times\VEClu{n}{\CORRII{\RMGamma}}{r}
            \label{EQdefThinShellHtangJump}
    \end{align}
where the indexes 1 and 2 refer to the two sides of the surface $\SCAlu{\RMGamma}{\RM{H}}{r}$, \CORRII{and the magnetic permeability is assumed continuous, due to the insulation of the source domains.} At the same time, the tangential component of $\VEC{E}$ and the normal component of $\VEC{B}$ are continuous} \CORRII{across $\SCAlu{\RMGamma}{\RM{H}}{r}$. These properties are used in the next seciton to determine the interface conditions for the fields within $\SCAlu{\RMGamma}{\RM{H}}{r}$ and in $\SCAl{\RMOmega}{\RM{A}}$, respectively.}


\subsection{Discretization}
    \label{SUBSEC_ThinShellDiscretization}

\CORRI{The discrete field problem is derived with respect to the general weak formulation~\eqref{EQweakAampereMaxwell}-\eqref{EQweakTheatBalance}. \CORRII{First,} the geometry of the shell is taken into account by transforming the volume integrals in $\SCAlu{\RMOmega}{\RM{H}}{r}$ into surface integrals in $\SCAlu{\RMGamma}{\RM{H}}{r}$, weighted with the thickness of the tape $\SCAlu{\RMdelta}{\RM{t}}{r}$.} \CORRII{Next,} the field properties \CORRII{at the shell boundary} are taken into account. In detail, the \CORRII{third} integral in~\eqref{EQweakAampereMaxwell} \CORRII{is modified with~\eqref{EQdefThinShellCurrentK} and~\eqref{EQdefThinShellHtangJump}, providing the field interface for $\SCAl{\RMOmega}{\RM{A}}$}. The first integral in~\eqref{EQweakHfaraday} vanishes, due to tangential continuity of $\VEC{E}$. The second integral in~\eqref{EQweakHfaraday}, reduced into a surface integral, \CORRII{is modified with~\eqref{EQinterfaceAHnormB}, providing the field interface for $\SCAlu{\RMGamma}{\RM{H}}{r}$} via the normal continuity of $\VEC{B}$. \CORRII{Finally}, the field constraints~\eqref{EqThinShellConstaintJ} and~\eqref{EqThinShellConstaintH} are included by defining a suitable set of edge functions $\VEClu{v}{q}{r}$ \CORRII{which discretize} the field $\VECu{H}{r}$ as
    \begin{align}
        \EQdefHThinShelldiscrete,
        \label{EQdefHThinShelldiscrete}
    \end{align}
with $\SCAlu{N}{q}{r}$ representing the nodal basis functions defined at the surface of the \CORRI{thin shell}, and zero elsewhere. \CORRII{Similar nodal functions are also used for discretizing the voltage distribution function $\VECu{\ITchi}{r}$, as .} 

Once all the field variables are replaced with their discrete counterparts and boundary conditions are considered, the following discrete problem is obtained:
    \begin{align}
        \!\!\!\!\!\!\EQmatAHNORMTdiscrete.
        \label{EQmatAHNORMTdiscrete}
    \end{align}
The coefficients in~\eqref{EQmatAHNORMTdiscrete} differing from those in~\eqref{EQmatAHTdiscrete} are given for $r=1,...,\SCAl{N}{\RM{r}}$ as
    \begin{align} 
        \EQdiscreteCoeffAHNORMlabelQ,    
            \label{EQdiscreteCoeffAHNORMlabelQ}       \\
        \EQdiscreteCoeffAHNORMlabelKrho, 
            \label{EQdiscreteCoeffAHNORMlabelKrho} \\
                    \EQdiscreteCoeffAHNORMlabelCHI,  
            \label{EQdiscreteCoeffAHNORMlabelCHI} \\
                    \EQdiscreteCoeffAHNORMlabelPJ.  
            \label{EQdiscreteCoeffAHNORMlabelPJ} 
    \end{align}
The elements composing the heat source $\MAT{\BAR{q}}(\cdot)$ in the right-hand side of~\eqref{EQmatAHNORMTdiscrete} are given by the sum of~\eqref{EQdiscreteCoeffTlabelPjoule} and~\eqref{EQdiscreteCoeffAHNORMlabelPJ}, the second term occurring only in the \CORRI{thin shell}. 


\subsection{Current Sharing}
    \label{SUBSEC_CurrentSharing}
    
The layered structure of the tape allows the transport current to redistribute between all the conducting layers. \CORRII{Such} current sharing regime occurs when the resistivity in~\eqref{EQdefJpowerLaw} suddenly increases. As example, this can be caused by a temperature increase \CORRI{that reduces} the critical current density. The \CORRI{thin shell} approximation \CORRII{needs to} account for the current sharing regime and the \CORRII{consequent} Joule losses, which serve as input for the heat balance equation. 

The electrical behavior of the \CORRI{thin shell} is represented by a parallel connection of two \CORRI{conducting} paths. One path is associated to the superconducting layer, the other to a bulk homogenization of the normal conducting layers. 
\begin{figure}[tb]
  \centering
	\includegraphics[width=8.0cm]{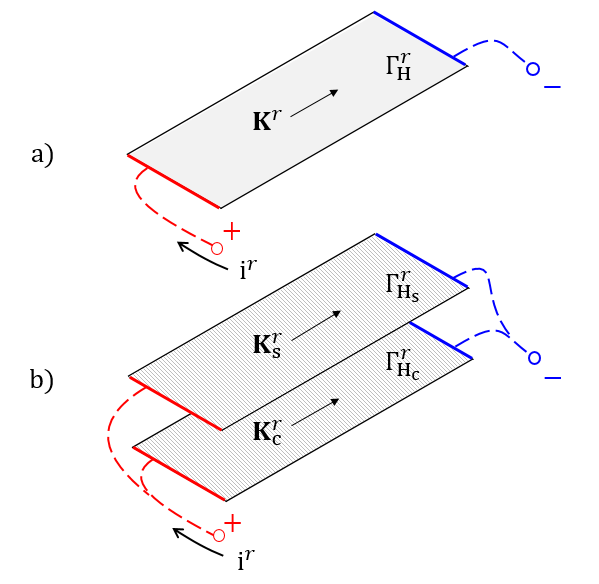}
	\caption{Equivalent electrical behavior of the \CORRI{thin shell} approximation. The surface $\SCAlu{\RMGamma}{\RM{H}}{r}$ in a) is equivalent to b), where the two conductive paths $\SCAlu{\RMGamma}{\RM{H_s}}{r}$ and $\SCAlu{\RMGamma}{\RM{H_c}}{r}$ represent the superconducting layer and the normal conducting layer, respectively.}
	\label{FIG04_ThinShellCurrentSharing}
\end{figure}
The \CORRI{thin shell} carries the surface current density $\VECu{K}{r}$ defined in~\eqref{EQdefThinShellCurrentK}, as shown in Fig.~\ref{FIG04_ThinShellCurrentSharing}a. At the same time, the superconducting and \CORRII{normal} conducting layers are modeled by two perfectly superimposed \CORRI{thin shells} $\SCAlu{\RMGamma}{\RM{H_s}}{r}$ and $\SCAlu{\RMGamma}{\RM{H_c}}{r}$ (see Fig.~\ref{FIG04_ThinShellCurrentSharing}b), with surface resistivities $\SCAlu{\ITrho}{\RMGamma_{\RM{s}}}{r}$ and $\SCAlu{\ITrho}{\RMGamma_{\RM{c}}}{r}$, and surface currents $\VEClu{K}{\RM{s}}{r}$ and $\VEClu{K}{\RM{c}}{r}$, respectively. The geometrical construction of the \CORRII{shells} enforces $\SCAlu{\RMGamma}{\RM{H}}{r}=\SCAlu{\RMGamma}{\RM{H_s}}{r}=\SCAlu{\RMGamma}{\RM{H_c}}{r}$. Thus, Kirchhoff's current (KCL) and voltage (KVL) laws hold true also in their differential formulation. The KCL allows to introduce the current sharing index $\SCAu{\ITlambda}{r} \in [0,1]$. The index splits $\VECu{K}{r}$ into its superconducting and normal conducting components, as 
    \begin{align}
        \VEClu{K}{\RM{s}}{r} = \SCAu{\ITlambda}{r}\VECu{K}{r}\ \RM{and} \      
        \VEClu{K}{\RM{c}}{r} = (1-\SCAu{\ITlambda}{r})\VECu{K}{r}.
        \end{align}
The KVL allows to \CORRII{formulate} the electric field balance of the superconducting and normal conducting layers as
    \begin{align}
        \eqCurrentSharingKVL.
        \label{eqCurrentSharingKVL}
        \end{align}
Using for $\SCAlu{\ITrho}{\RMGamma_{\RM{s}}}{r}$ the definition in~\eqref{EQdefJpowerLaw}, the surface resistivities in~\eqref{eqCurrentSharingKVL} are given as
    \begin{align}
        \eqCurrentSharingRhoScSurf \label{eqCurrentSharingRhoScSurf},
        \ \EQdefCriticalCurrentIndex \\[1.0ex]
        \eqCurrentSharingRhoNcSurf, \label{eqCurrentSharingRhoNcSurf}
        \end{align}
where $\CORRII{\SCAlu{K}{\RM{crit}}{r}}=\SCAlu{\RMdelta}{\RM{t_s}}{r}\SCAl{J}{\RM{crit}}$ is the critical surface current density, $\SCAu{k}{r}$ is defined as the surface current density saturation index, and $\SCAl{N}{\RM{c}}$ is the number of \CORRII{materials composing} the normal conducting layer of the tape. Substituting~\eqref{eqCurrentSharingRhoScSurf} and~\eqref{eqCurrentSharingRhoNcSurf} into~\eqref{eqCurrentSharingKVL} leads to the following root-finding problem: find $f(\SCAu{\ITlambda}{r})=0$, with
    \begin{align}
        \eqCurrentSharingLambda,\ \ \eqCurrentSharingAlpha\!\!\!\!\!\!\geq0.
        \label{eqCurrentSharingLambda}
    \end{align}
\begin{figure}[tb]
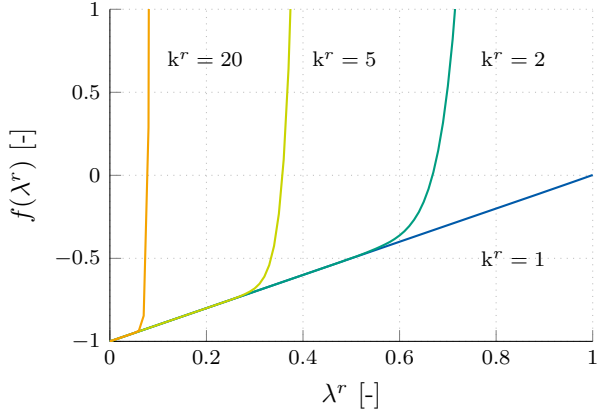

  \centering
	\FigureCurrentSharingFunction
	\caption{Behavior of the polynomial associated to the zero-finding problem, as a function of the current sharing index $\SCAu{\ITlambda}{r}$ in the superconducting layer. The curves are parametrized by the surface current density saturation index $\SCAu{k}{r}$ in the tape.}
	\label{FIG05_CurrentSharingFunction}
\end{figure}
Such equation cannot be solved analytically, due to the current dependency in $\SCAlu{\ITrho}{\RMGamma_{\RM{s}}}{r}$. The continuity property of the polynomial and the intermediate zero theorem ensure that for $\SCAu{\ITalpha}{r}>0$, $f(\SCAu{\ITlambda}{r})=0$ admits at least one real root in the interval $[0,1]$. Moreover, applying Descartes' rule of signs~\cite{collins1976} to the polynomial $f(\SCAu{\ITlambda}{r})$ guarantees the existence of only one real and positive root, at most. 

It is worth observing that the behavior of the polynomial derivative $f'(\SCAu{\ITlambda}{r})$ strongly depends \CORRI{on} the current regime in the tape. For small currents ($\SCAu{k}{r}\to0$) $f'(\SCAu{\ITlambda}{r})=0$, \CORRII{\CORRII{whereas}} for currents beyond the critical current of the tape (${\SCAu{k}{r}\to+\infty}$)  $f'(\SCAu{\ITlambda}{r})=+\infty$, as shown in Fig.~\ref{FIG05_CurrentSharingFunction}. Due to the behavior of $f'(\SCAu{\ITlambda}{r})$, the Newton-Raphson method is abandoned in favor of the bisection method, which guarantees linear convergence. Once $\SCAu{\ITlambda}{r}$ is determined, \CORRII{it is used in}~\eqref{eqCurrentSharingRhoScSurf} and the equivalent surface resistivity {$\SCAlu{\ITrho}{\CORRII{\RMGamma}_{\RM{eq}}}{r}$} \CORRII{is} calculated as 
    \begin{align}
        \eqCurrentSharingRhoEq
        \label{eqCurrentSharingRhoEq}
    \end{align}
The current sharing algorithm was implemented as an inner loop within the time-stepping algorithm.

\section{Numerical Results} 
    \label{SEC_NumericalResults}
    
\CORRI{The results section has two primary objectives: 1) to verify and validate the proposed formulation, and 2) to show that the numerical implementation is sufficiently flexible for simulating full size applications.} \CORRI{For the first objective, the formulation is validated by} \CORRII{comparison with available reference results}. In detail, the full set of equations proposed in~\eqref{EQmatAHTdiscrete} is cross-checked with \CORRI{a 3D benchmark model proposed in}~\cite{kapolka2018three}, and the \CORRI{thin shell} approximation proposed in~\eqref{EQmatAHNORMTdiscrete} is cross-checked with \CORRI{a 2D benchmark model proposed in}~\cite{rodriguez2011towards}. The \CORRII{benchmark} models are based on the $\VEC{H}$ formulation~\cite{bossavit1982mixed}, solving for the magnetic field strength vector. Both the \CORRII{benchmark} models are available at~\cite{HTS2019website}. 
\CORRI{For the second objective, the formulation is applied to the 2D time-domain simulation of the \FEATHER magnet. In detail, the current-voltage behavior and the electrodynamics of the coil is presented.}

The coupled $\VEC{A}\HYPHEN\VEC{H}$ field formulation was implemented in the proprietary {FEM} solver {COMSOL} Multiphysics$^{\circledR}$~\cite{comsol2005comsol}. \CORRI{The set of magneto-thermal equations are solved monolithically, using an adaptive time-stepping method based on the fifth-order {BDF}.} All the simulations were carried out on a standard workstation (Intel Xeon CPU E5-2667 v4 @ 3.40GHz, 128 GB of RAM, Windows 7 operative system).


\subsection{{3D} Superconducting Bulk}
    \label{SUBSEC_3DSuperconductingBulk}
    
\begin{figure}[tb]
  \centering
	\includegraphics[width=8.0cm]{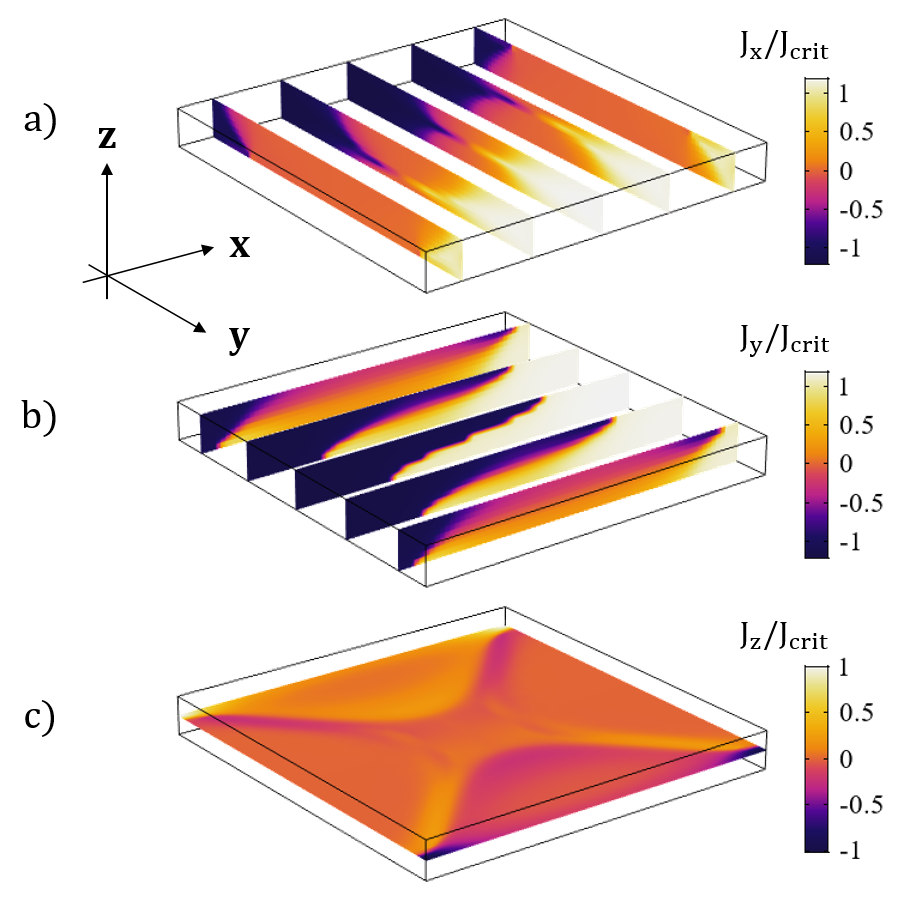}
	\caption{Current density in the superconducting bulk, normalized \CORRII{with} respect to the critical current density, \CORRII{and} rendered by means of cut planes. The normalized current density components are given respect to a)
	$\SCAl{J}{\RM{x}}$, b) $\SCAl{J}{\RM{y}}$ and c) $\SCAl{J}{\RM{z}}$.}
	\label{FIG06_Block3DCurrentDensity}
\end{figure}
The $\VEC{A}\HYPHEN\VEC{H}$ formulation is used to analyze the currents induced in a superconducting rectangular bulk exposed to a time-varying magnetic field. The details of the model are discussed in~\cite{kapolka2018three}. The parameters for the superconducting material are chosen as $\CORRII{\SCAl{J}{\RM{crit}}}=\CORRI{\SI{1e+8}{\ampere\per\meter\squared}}$ and $\SCA{n}=25$. The orientation of the source field is such that the currents induced in the superconducting bulk show components in all the three dimensions. \CORRII{For this reason,} only a {3D} model can capture the behavior of the induced currents. The superconducting bulk was discretized by means of a structured mesh. Both a coarse and a fine resolution scheme were used for the mesh generation, leading to $24\times24\times8$ and $71\times71\times7$ cell elements, respectively. Fig.~\ref{FIG06_Block3DCurrentDensity} shows the three spatial components of the current density distribution in the bulk. The distribution is consistent with the reference work, leading to a Joule dissipation of \CORRII{$\SI{4.7}{\milli\joule}$/cycle}, which is in \CORRII{$\SI{1}{\%}$} agreement with respect to the reference. The obtained computational time was 0.2 and 2.2 hours for the coarse and the fine mesh, respectively. 
This specific application requires a computational time which is comparable to the one \CORRII{obtained by using a monolithic $\VEC{H}$ formulation.}


\subsection{{2D} Superconducting Tape}
    \label{SUBSEC_2DSuperconductingTape}
    
\begin{figure}[tb]
  \centering
	\includegraphics[width=8.0cm]{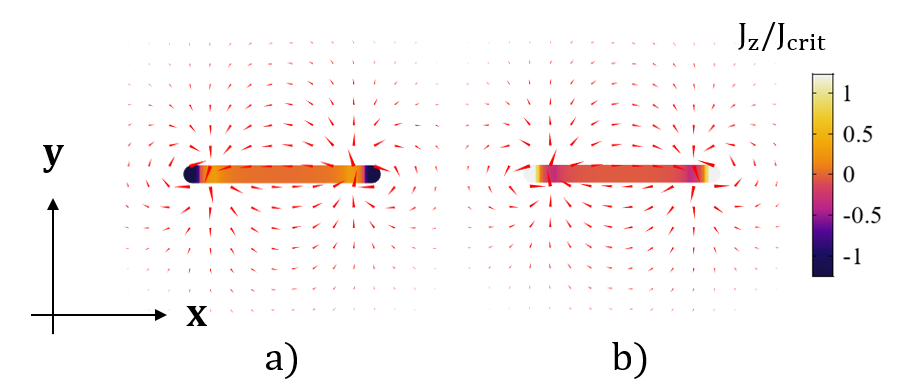}
	\caption{Normalized current density distribution in the superconducting tape for a sinusoidal source current of $\SI{100}{\ampere}$ and $f=\SI{100}{\hertz}$, at a) $t=\SI{5}{\milli\second}$ and b) $t=\SI{10}{\milli\second}$. The colormap gives the normalized current density, with respect to the critical current density. The cones represent the magnetic flux density distribution.}
	\label{FIG07_Tape2DCurrentDensity}
\end{figure}
\begin{figure}[tb]
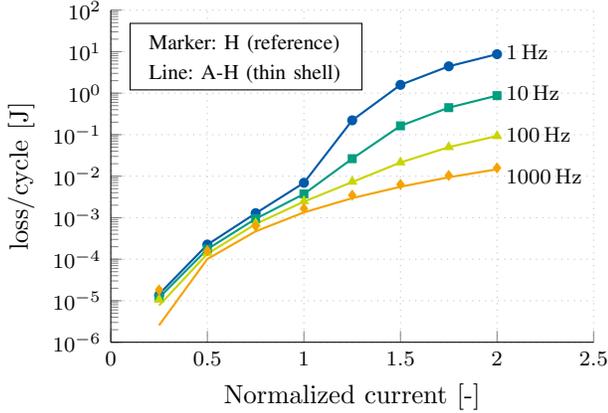

  \centering
	\TapeTwoDCurrentDensity
	\caption{Joule losses per cycle in a single tape powered with a sinusoidal current. The losses are given for both, the reference model (markers) and the $\VEC{A}\HYPHEN\VEC{H}$ formulation (lines). The horizontal axis, representing the source current, is normalized with respect to the tape critical current of $\SI{200}{\ampere}$. The losses are determined as function of the frequency of the source current, and are given at $1$, $10$, $100$, $\SI{1000}{\hertz}$. A monotonic increase is observed for all the curves.
	}
	\label{FIG08_AClosses}
\end{figure}
\begin{figure}[tb]
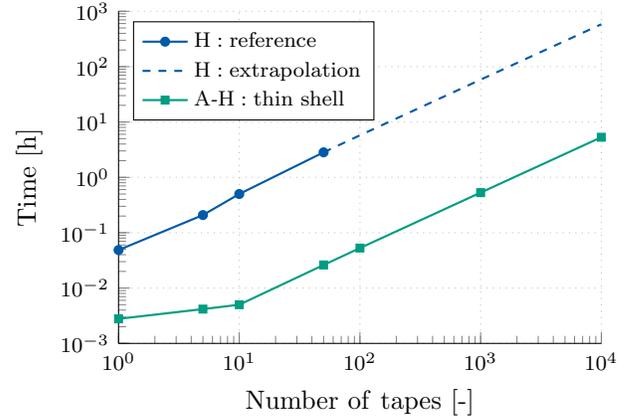

\centering
    \FigComparisonComputationalTime
    \caption{Computational time in hours, as function of the number of tapes included in the model. The reference model features 2D domains for the superconducting tapes, \CORRII{whereas} the $\VEC{A}\HYPHEN\VEC{H}$ formulation relies on the \CORRI{thin shell} approximation.}
    \label{FIG09_ComputationalTime}
\end{figure}

The $\VEC{A}\HYPHEN\VEC{H}$ formulation is applied in combination with the \CORRI{thin shell} approximation to an individual HTS tape in a transverse-field {2D} configuration (Fig.~\ref{FIG07_Tape2DCurrentDensity}). \CORRI{The model accounts for the current-sharing regime by including the normal conducting part of the tape as discussed in Section~\ref{SUBSEC_CurrentSharing}.}

\CORRI{The parameters related to the superconducting part of the tape} are set to $\CORRII{\SCAl{J}{\RM{crit}}}=\CORRI{\SI{5e+10}{\ampere\per\meter\squared}}$ and $\SCA{n}=20$, \CORRI{whereas a constant resistivity $\SCA{\ITrho}=\SI{1e-10}{\ohm\per\meter}$ is used for the normal conducting part}, \CORRI{accordingly to the parameters} \CORRII{chosen} in the reference model. The Joule losses are shown in Fig.~\ref{FIG08_AClosses}) and compared against the reference model, which used explicit 2D domains for the superconducting tapes. The tape is powered with a sinusoidal current $\SCAl{i}{\RM{s}}{}(t)=\SCAl{I}{\RM{0}}\RM{sin}(\CORRII{2\pi f} t)$. Currents with frequencies $1$, $10$, $100$, $\SI{1000}{\hertz}$ are applied as source. The current amplitude $\SCAl{I}{\RM{0}}$ ranges between $0.25\SCAl{I}{\CORRII{\RM{crit}}}$ and $2\SCAl{I}{\CORRII{\RM{crit}}}$, where $\SCAl{I}{\CORRII{\RM{crit}}}=200$ A is the critical current of the tape. By increasing the frequency and the magnitude of the applied current, it is possible \CORRII{to quantify the approximations introduced by the thin shell geometry and the equivalent surface resistivity, respectively.}

The Joule losses per cycle are reported in Fig.~\ref{FIG08_AClosses}. \CORRI{For \CORRII{sub-critical} currents ($\SCA{I}<\SCAl{I}{\CORRII{\RM{crit}}}$), the difference of the curves decreases but does not vanish. The difference is due to a weak frequency dependency in the losses per cycle (see for example~\cite{sander2010fem}). \CORRII{For over-critical currents} ($\SCA{I}>\SCAl{I}{\CORRII{\RM{crit}}}$), the losses are \CORRII{dominated by the contribution from} the normal conducting part of the tape. This is reflected by the substantial frequency dependence in the curves.} The agreement \CORRI{between the reference model and the coupled field formulation} is satisfactory within the \CORRI{proposed} parameter space and \CORRI{it shall be}  sufficient for simulating the typical magneto-thermal transients occurring in accelerator magnets. 

The influence of the \CORRI{thin shell} approximation on the computational time of the model is shown in Fig.~\ref{FIG09_ComputationalTime}. In detail, the same equations have been solved for models featuring an increasing number of tapes, stacked on each other, for both the $\VEC{H}$ and the $\VEC{A}\HYPHEN\VEC{H}$ formulations. The performance improvement achieved with the \CORRI{thin shell} approximation is about two orders of magnitude. \CORRI{For the most complex model, which contains $10^4$ tapes, a computational time of about 8 hours was registered.}


\subsection{\FEATHER Magnet}
    \label{SUBSEC_FeatherM2Magnet}

The \FEATHER magnet~\cite{van2018powering} is simulated in a 2D planar field configuration. The coil of the magnet is made of a Roebel cable~\cite{goldacker2007roebel},~\CORRII{\cite{goldacker2014roebel}}, \CORRI{whose} dynamics occurring in the fully transposed tapes cannot be completely represented by a 2D model. For that reason, the coil dynamics is approximated by assuming a homogeneous current redistribution along the tapes in the cable. Such approximation does not take into account 3D localized phenomena (e.g., a local temperature increase) which are not considered in the numerical analysis.

\CORRI{Because of the high} number of superconducting domains in the cross-section of the magnet (648 tapes over the four quadrants), the \CORRI{thin shell} approximation proposed in Section~\ref{SEC_ThinShellApproximation} is applied. In detail, the magnet model is obtained by implementing the \CORRI{formulation} proposed in~\eqref{EQmatAHNORMTdiscrete}. With regards to the tape critical current density \CORRII{$\SCAl{J}{\RM{crit}}$}, a field-angle dependent relation is used~\cite{fleiter2014characterization}, with fitting parameters provided by the producer of the tape~\cite{SUNAM2019website}. The power law index is chosen as $\SCA{n}=20$. 

The magnet is powered by imposing an external current to the superconducting coil, following the cycle shown in Fig.~\ref{FIG10_FeatherM2Powering}. The powering of the coil occurs in two steps, a pre-cycle and a ramp, with a rate of \CORRII{$\SI{5}{\ampere\per\second}$}. The pre-cycle follows a trapezoidal profile, bringing the current from $0$ to $\SI{5}{\kilo\ampere}$, then back to $\SI{0.1}{\kilo\ampere}$. The consecutive ramp brings the magnet back to $\SI{5}{\kilo\ampere}$, which is then kept constant. The voltage drop calculated across the coil is also given in Fig.~\ref{FIG10_FeatherM2Powering}. The saturation due to the magnetization of the iron yoke and the consequent reduction of the differential inductance of the coil are clearly visible. 

Fig.~\ref{FIG11_FeatherCoilM2CurrentDensity} shows the normalized current density distribution in the coil, for a current of $\SI{3}{\kilo\ampere}$ reached during: a) the positive slope of the pre-cycle, b) the negative slope of the pre-cycle, c) the ramp. The current density in Fig.~\ref{FIG11_FeatherCoilM2CurrentDensity}c shows along the tapes two discontinuities in the polarity. This might be justified by considering the induced eddy currents as a persistent phenomenon, showing a decay time longer than the duty cycle of the coil. Thus, for each dynamic phase of the powering cycle, new eddy currents are induced at the edges of the tapes, pushing inward and screening any previous dynamic effect. Such behavior is compatible with the critical state model~\cite{bean1964magnetization}, differing by showing a finite decay time for the eddy currents. The overall computational time was about 2 hours.
\begin{figure}[tb]
  \centering
	\FigPoweringFeather
	\caption{Voltage and current profiles applied to the \FEATHER magnet. The powering of the coil occurs in two steps, a pre-cycle and a ramp, with a rate of \CORRII{$\SI{5}{\ampere\per\second}$}. The pre-cycle follows a trapezoidal profile, bringing the current to the nominal value of $\SI{5}{\kilo\ampere}$ and then to $\SI{0.1}{\kilo\ampere}$. The ramp brings the magnet back to the nominal current. }
	\label{FIG10_FeatherM2Powering}
\end{figure}
\begin{figure}[tb]
  \centering
	\includegraphics[width=8.0cm]{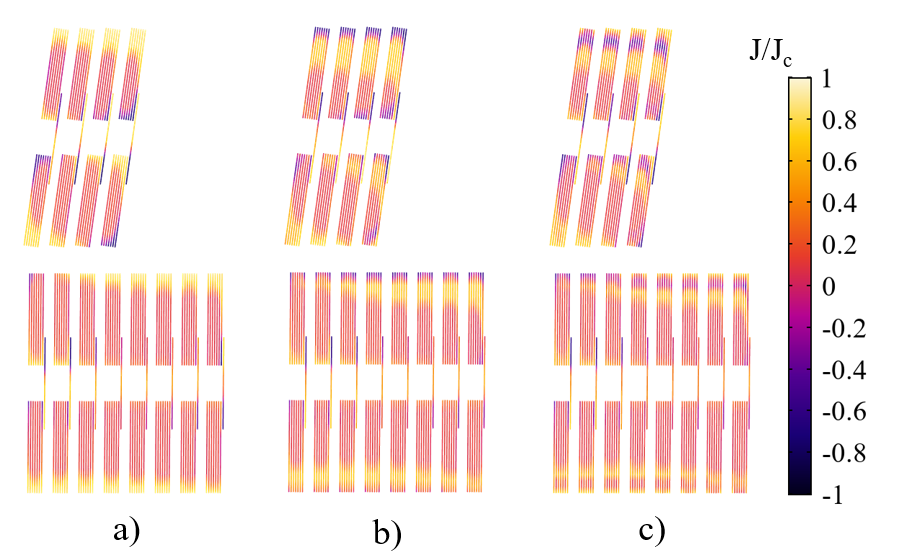}
	\caption{Normalized current density distribution in the \FEATHER coil,  for a current of $\SI{3}{\kilo\ampere}$ reached during: a) the positive slope of the pre-cycle, b) the negative slope of the pre-cycle, c) the ramp.}
	\label{FIG11_FeatherCoilM2CurrentDensity}
\end{figure}


\section{Conclusion and Outlook} 
    \label{SEC_ConclusionOutlook}
    
Numerical simulations play a \CORRI{decisive} role in supporting the technology switch towards HTS magnets for particle accelerators. This paper proposes a coupled $\VEC{A}\HYPHEN\VEC{H}$  formulation for the time-domain simulation of devices containing high-temperature superconducting tapes and cables. The formulation enables to consider the power law modeling the superconducting material behavior. The general 3D case, as well as a \CORRI{thin shell} approximation in which the thickness of thin superconducting layers is not resolved by the computational mesh are presented. \CORRII{Moreover, a field-circuit coupling interface suitable for co-simulation is derived.} The experiments show that the overall method is numerically stable for $\SCA{\ITsigma}\to\infty$ and allows to simulate superconducting devices with thousands of tapes with an affordable computation time. For the first time, the dynamic behavior of the high-temperature superconducting cable with the \FEATHER magnet could be calculated with a sufficient resolution in space and time. The formulation allows to quantify the magnetic field quality and the thermal behavior in HTS-based devices, and will be applied for the design of future HTS accelerator magnets.    
    


\section{Acknowledgements} 
    \label{Acknowledgements}

The authors would like to acknowledge the fruitful collaboration between CERN and the Technische Universit{\"a}t Darmstadt, within the framework of the STEAM project~\cite{STEAM2019website}. The authors would like to thank \CORRII{D. Martins Araujo for fruitful discussions on the mathematical details of the formulation,} and S. Friedel and the team of COMSOL Switzerland for continued support during the numerical implementation of the formulation.  


\bibliographystyle{IEEEtran}
    \bibliography{  Bib/BIB_IEEEabrv,
                    Bib/BIB_article,
                    Bib/BIB_book,
                    Bib/BIB_misc,
                    Bib/BIB_thesis
                    }
   
              
\end{document}